\documentclass{appolb}
\usepackage[dvips]{graphicx}
\usepackage[dvips]{color}
\usepackage{hyperref,breakurl,amsmath,amssymb,pst-node,eepic,tipa}

\newcommand{\ieNotAPPB}{\emph{i.e.}, }
\newcommand{\ii}{\mathrm{i}}
\newcommand{\ee}{\mathrm{e}}
\newcommand{\dd}{\mathrm{d}}
\newcommand{\DD}{\mathrm{D}}
\newcommand{\re}{\mathrm{Re}}
\newcommand{\im}{\mathrm{Im}}
\newcommand{\bTr}{\mathrm{bTr}}
\newcommand{\Det}{\mathrm{Det}}
\newcommand{\Id}{\mathbf{1}}
\newcommand{\Zero}{\mathbf{0}}

\newcommand{\la}{\left\langle}
\newcommand{\ra}{\right\rangle}
\newcommand{\tot}{\textrm{tot.}}

\begin{document}

\title{Eigenvalues and Singular Values\\of Products of Rectangular Gaussian Random Matrices\\(The Extended Version)\thanks{Presented by Giacomo Livan at the 23rd Marian Smoluchowski Symposium on Statistical Physics ``Random Matrices, Statistical Physics and Information Theory,'' September 26--30, 2010, Krak\'{o}w, Poland.}}

\author{Zdzislaw Burda
\and
Maciej A. Nowak
\address{The Marian Smoluchowski Institute of Physics, Jagiellonian University, Reymonta 4, 30--059 Krak\'{o}w, Poland}
\address{The Mark Kac Complex Systems Research Centre, Jagiellonian University, Reymonta 4, 30--059 Krak\'{o}w, Poland}
\and
Andrzej Jarosz
\address{The Henryk Niewodnicza\'{n}ski Institute of Nuclear Physics, Polish Academy of Sciences, Radzikowskiego 152, 31--342 Krak\'{o}w, Poland}
\and
Giacomo Livan
\thanks{\href{mailto:giacomo.livan@pv.infn.it}{{\tt giacomo.livan@pv.infn.it}} (corresponding author)}
\address{Dipartimento di Fisica Nucleare e Teorica, Universit\`{a} degli Studi di Pavia, Via Bassi 6, 27100 Pavia, Italy}
\address{Istituto Nazionale di Fisica Nucleare, Sezione di Pavia, Via Bassi 6, 27100 Pavia, Italy}
\and
Artur Swiech
\address{The Marian Smoluchowski Institute of Physics, Jagiellonian University, Reymonta 4, 30--059 Krak\'{o}w, Poland}
}

\maketitle

\begin{abstract}
This is a longer version of our article~\cite{BurdaJaroszLivanNowakSwiech2010}, containing more detailed explanations and providing pedagogical introductions to the methods we use.

We consider a product of an arbitrary number of independent rectangular Gaussian random matrices. We derive the mean densities of its eigenvalues and singular values in the thermodynamic limit, eventually verified numerically. These densities are encoded in the form of the so--called $M$--transforms, for which polynomial equations are found. We exploit the methods of planar diagrammatics, enhanced to the non--Hermitian case, and free random variables, respectively; both are described in the appendices. As particular results of these two main equations, we find the singular behavior of the spectral densities near zero. Moreover, we propose a finite--size form of the spectral density of the product close to the border of its eigenvalues' domain. Also, led by the striking similarity between the two main equations, we put forward a conjecture about a simple relationship between the eigenvalues and singular values of any non--Hermitian random matrix whose spectrum exhibits rotational symmetry around zero.
\end{abstract}



\section{Introduction and the Main Results}
\label{s:IntroductionAndTheMainResults}


\subsection{Introduction}
\label{ss:Introduction}


\subsubsection{Products of Random Matrices}
\label{sss:ProductsOfRandomMatrices}

The problem of \emph{multiplication of random matrices} received a considerable measure of attention from the random matrix theory (RMT) community recently~\cite{CrisantiPaladinVulpiani1993,JacksonLautrupJohansenNielsen2002,Caswell2000,GredeskulFreilikher1990,TulinoVerdu2004,Beenakker1997,NarayananNeuberger2007,BlaizotNowak2008,GudowskaNowakJanikJurkiewiczNowak2003,LohmayerNeubergerWettig2008,BanicaBelinschiCapitaineCollins2007,BenaychGeorges2008}. In this matter, one is interested in studying the statistical properties (in particular, the average distribution of the eigenvalues) of the product
\begin{equation}\label{eq:ProductDefinition}
\mathbf{P} \equiv \mathbf{A}_{1} \mathbf{A}_{2} \ldots \mathbf{A}_{L}
\end{equation}
of $L \geq 1$ random matrices \smash{$\mathbf{A}_{l}$}, $l = 1 , 2 , \ldots , L$. In the basic version of the problem, all these matrices are assumed to be statistically independent, to which condition we too will adhere.

In order for these matrices to be multiplicable, the dimensions of each \smash{$\mathbf{A}_{l}$} have most generally to be of the form \smash{$N_{l} \times N_{l + 1}$}, where \smash{$N_{1} , N_{2} , \ldots , N_{L} , N_{L + 1}$} are integers. Then, $\mathbf{P}$ has sizes \smash{$N_{1} \times N_{L + 1}$}; if this matrix is to have eigenvalues, it must obviously be square, \ieNotAPPB \smash{$N_{L + 1} = N_{1}$}.

Moreover, in RMT one typically considers the ``thermodynamic limit'' in which the matrices are infinitely large, but their dimensions have the same order of magnitude, \ieNotAPPB their ratios stay finite; here, we decide to define these ratios w.r.t. \smash{$N_{L + 1}$},
\begin{equation}\label{eq:ThermodynamicLimit}
N_{l} \to \infty , \quad \textrm{with} \quad R_{l} \equiv \frac{N_{l}}{N_{L + 1}} = \textrm{finite} , \quad \textrm{for} \quad l = 1 , 2 , \ldots , L , L + 1 .
\end{equation}
(Obviously, \smash{$R_{L + 1} = 1$}; if we additionally take \smash{$N_{L + 1} = N_{1}$}, then there also is \smash{$R_{1} = 1$}.) In this limit, all finite--size corrections are lost, such as universal oscillations of the spectrum at its edges.


\subsubsection{Products of Square Girko--Ginibre Random Matrices}
\label{sss:ProductsOfSquareGirkoGinibreRandomMatrices}

To date, only various ensembles of \emph{square} (\ieNotAPPB \smash{$N_{1} = N_{2} = \ldots = N_{L + 1} \equiv N$}) random matrices have been investigated in the context of the multiplication task. For instance, the authors of~\cite{BurdaJanikWaclaw2010} considered each \smash{$\mathbf{A}_{l}$} to consists of complex entries whose real and imaginary parts are all IID Gaussian random variables with zero mean and variance \smash{$\sigma_{l}^{2} / 2 N$}; in other words, such a random matrix model, called the ``Girko--Ginibre ensemble''~\cite{Ginibre1965,Girko1984,Girko1985}, has the probability measure,
\begin{equation}\label{eq:SquareGGMeasure}
\dd \mu \left( \mathbf{A}_{l} \right) \propto \ee^{- \frac{N}{\sigma_{l}^{2}} \Tr \left( \mathbf{A}_{l}^{\dagger} \mathbf{A}_{l} \right)} \DD \mathbf{A}_{l} ,
\end{equation}
where the flat measure \smash{$\DD \mathbf{A}_{l} \equiv \prod_{a , b = 1}^{N} \dd ( \re [ \mathbf{A}_{l} ]_{a b} ) \dd ( \im [ \mathbf{A}_{l} ]_{a b} )$}, and the normalization constant has been omitted. Using the method of \emph{planar diagrams} and \emph{Dyson--Schwinger's equations}, which is the same technique we will employ here in section~\ref{s:TheEigenvaluesOfAProductOfRectangularGaussianRandomMatrices} and pedagogically describe in appendix~\ref{s:TheDiagrammaticApproachToSolvingMatrixModels}, one may derive (\S\ref{sss:TheNonHolomorphicGreensFunctionForTheGirkoGinibreEnsembleFromPlanarDiagrams}) that on average, the eigenvalues of a Girko--Ginibre matrix are scattered within a centered circle of radius \smash{$\sigma_{l}$} with the uniform density,
\begin{equation}\label{eq:GGMeanSpectralDensity}
\rho_{\mathbf{A}_{l}} ( \lambda , \overline{\lambda} ) = \left\{ \begin{array}{ll} \frac{1}{\pi \sigma_{l}^{2}} , & \quad \textrm{for} \quad | \lambda | \leq \sigma_{l} , \\ 0 , & \quad \textrm{for} \quad | \lambda | > \sigma_{l} . \end{array} \right.
\end{equation}

Employing non--Hermitian planar diagrammatics again, yet in a more demanding situation, the authors of~\cite{BurdaJanikWaclaw2010} (see their equation (5)) discovered that the eigenvalues of the product $\mathbf{P}$ (\ref{eq:ProductDefinition}) of such Girko--Ginibre matrices also fill a centered circle, of radius
\begin{equation}\label{eq:Sigma}
\sigma \equiv \sigma_{1} \sigma_{2} \ldots \sigma_{L} ,
\end{equation}
with the average density given by a surprisingly simple expression,
\begin{equation}\label{eq:SquareGGProductMeanSpectralDensity}
\rho_{\mathbf{P}} ( \lambda , \overline{\lambda} ) = \left\{ \begin{array}{ll} \frac{1}{L \pi \sigma^{2}} \left| \frac{\lambda}{\sigma} \right|^{- 2 \left( 1 - \frac{1}{L} \right)} , & \quad \textrm{for} \quad | \lambda | \leq \sigma , \\ 0 , & \quad \textrm{for} \quad | \lambda | > \sigma . \end{array} \right.
\end{equation}
Remarkably, this formula remains valid even if the constituent matrices are not identically distributed, just with the assumptions of independence and Gaussianity retained, \ieNotAPPB they may come from different Gaussian ensembles, such as GUE, GOE, or the so--called ``Gaussian elliptic ensembles.'' Moreover, it has been conjectured~\cite{BurdaJanikWaclaw2010} that (\ref{eq:SquareGGProductMeanSpectralDensity}) holds for an even wider class of matrices, such as ones having independent entries fulfilling the Pastur--Lindeberg condition --- in this sense, the result (\ref{eq:SquareGGProductMeanSpectralDensity}) is \emph{universal}. One unexpected implication of this universality is that a product of random matrices whose spectra do not necessarily display rotational symmetry has the eigenvalue distribution on the complex plane which does possess rotational symmetry (\ieNotAPPB the average density depends only on $| \lambda |$).


\subsubsection{Non--Hermitian Random Matrix Models}
\label{sss:NonHermitianRandomMatrixModels}

Exploration of products of matrix models brings us almost inevitably into the realm of \emph{non--Hermitian} random matrices. The most pronounced difference between them and their Hermitian cousins is that their eigenvalues are generically complex, while Hermitian spectra must be real. In the thermodynamic limit, the eigenvalues of non--Hermitian ensembles cover two--dimensional domains on the complex plane, contrary to one--dimensional cuts in the Hermitian case.

Much of the RMT machinery --- such as the saddle--point method, orthogonal polynomials, the Efetov's supersymmetric technique, or the diagrammatic expansion and the free random variables calculus (see for example~\cite{Forrester2010,AndersonGuionnetZeitouni2010,Valko2009,StephanovVerbaarschotWettig2005,Verbaarschot2004,Mehta2004,DiFrancesco2004,Fyodorov2004,EdelmanWinSpring2004,EdelmanWinFall2004,BoutetdeMonvelKhorunzhy2001,Eynard2000,GuhrMullerGroelingWeidenmuller1998,DiFrancescoGinspargZinnJustin1995} for reviews; these last two techniques are used in this paper) --- has been developed in the Hermitian context, and tailored for real spectra. The necessity of dealing with complex spectra demands enhancements of these methods. In appendix~\ref{s:TheDiagrammaticApproachToSolvingMatrixModels} we present a self--contained crash course on how the planar diagrammatics can be applied to find average spectra of non--Hermitian random ensembles~\cite{JanikNowakPappWambachZahed1997,JanikNowakPappZahed1997,JanikNowakPappZahed2001}.

Leaving thus all the details for later, let us just mention now that as the information about the average spectrum of a Hermitian matrix model $\mathbf{H}$ is encoded in the ``spectral density'' \smash{$\rho_{\mathbf{H}} ( \lambda )$} (\ref{eq:MeanSpectralDensityDefinition}) --- which is a function of a real variable, or equivalently in the ``Green's function'' \smash{$G_{\mathbf{H}} ( z )$} (\ref{eq:GreensFunctionDefinition}) --- which is a holomorphic function everywhere except the cuts where the eigenvalues lie, or equivalently in the so--called ``$M$--transform,'' \smash{$M_{\mathbf{H}} ( z ) \equiv z G_{\mathbf{H}} ( z ) - 1$} (\ref{eq:MTransformDefinition}), so for a non--Hermitian matrix $\mathbf{X}$, one exploits an analogous set of concepts, namely, the spectral density \smash{$\rho_{\mathbf{X}} ( \lambda , \overline{\lambda} )$} (\ref{eq:NonHermitianMeanSpectralDensityDefinition}) --- which is now a function of a complex argument, denoted here by $\lambda$ and \smash{$\overline{\lambda}$} as independent variables (this is the object appearing in (\ref{eq:GGMeanSpectralDensity}) and (\ref{eq:SquareGGProductMeanSpectralDensity})), the ``non--holomorphic Green's function,'' \smash{$G_{\mathbf{X}} ( z , \overline{z} )$} (\ref{eq:NonHolomorphicGreensFunctionDefinition}), as well as the ``non--holomorphic $M$--transform,'' \smash{$M_{\mathbf{X}} ( z , \overline{z} ) \equiv z G_{\mathbf{X}} ( z , \overline{z} ) - 1$} (\ref{eq:NonHolomorphicMTransformDefinition}). Presentation in appendix~\ref{s:TheDiagrammaticApproachToSolvingMatrixModels} of the diagrammatic method augmented to handle non--Hermitian products is one of the motivations of this article.


\subsubsection{Singular Values and Free Random Variables}
\label{sss:SingularValuesAndFreeRandomVariables}

Besides eigenvalues, another important characteristics of a matrix model $\mathbf{P}$ is given by its ``singular values,'' which are defined as the square roots of the (real) eigenvalues of the Hermitian matrix
\begin{equation}\label{eq:SingularValuesDefinition}
\mathbf{Q} \equiv \mathbf{P}^{\dagger} \mathbf{P} .
\end{equation}
We will investigate the singular values in section~\ref{s:TheSingularValuesOfAProductOfRectangularGaussianRandomMatrices}, using and thereby promoting an approach called the ``free random variables'' (FRV) calculus. We sketch its fundamental implications in appendix~\ref{s:FreeRandomVariablesInANutshell}. The idea will be to rewrite $\mathbf{Q}$, through cyclic permutations of the constituent matrices, as a product of certain Hermitian ensembles. Now, for a Hermitian product of independent (more precisely, ``free'' (\ref{eq:FreenessDefinition}) --- this is a generalized notion of statistical independence, suited for matrix probability theory) Hermitian matrices --- FRV provides a ``multiplication algorithm'': Given two Hermitian random matrices, \smash{$\mathbf{H}_{1}$} and \smash{$\mathbf{H}_{2}$} whose product is still Hermitian, and which are mutually free, one begins from finding their ``$N$--transforms'' (\ref{eq:NTransformDefinition}), defined as the functional inverses of the respective $M$--transforms, \smash{$N_{\mathbf{H}_{1 , 2}} ( M_{\mathbf{H}_{1 , 2}} ( z ) ) = M_{\mathbf{H}_{1 , 2}} ( N_{\mathbf{H}_{1 , 2}} ( z ) ) = z$}. Then, a basic FRV theorem claims that the $N$--transform of the product \smash{$\mathbf{H}_{1} \mathbf{H}_{2}$} is in fact the product of the two $N$--transforms, up to a simple prefactor, \smash{$N_{\mathbf{H}_{1} \mathbf{H}_{2}} ( z ) = ( z / ( z + 1 ) ) N_{\mathbf{H}_{1}} ( z ) N_{\mathbf{H}_{2}} ( z )$} (\ref{eq:NonCommutativeMultiplicationLaw}). Inverting functionally the result, one obtains the $M$--transform, and thus the entire spectral information about \smash{$\mathbf{H}_{1} \mathbf{H}_{2}$}.


\subsection{The Main Results}
\label{ss:TheMainResults}


\subsubsection{The Aim of the Paper}
\label{sss:TheAimOfThePaper}

In this paper, our objective is to generalize the work of~\cite{BurdaJanikWaclaw2010} by calculating both the eigenvalues and singular values of a product (\ref{eq:ProductDefinition}) of \emph{rectangular} matrices with complex entries being IID Gaussian random variables. More precisely, for our $L$ random matrices \smash{$\mathbf{A}_{l}$} we will now allow arbitrary rectangularity ratios \smash{$R_{l}$} (\ref{eq:ThermodynamicLimit}), while the assumption of the real and imaginary parts of all the matrix elements of each \smash{$\mathbf{A}_{l}$} being IID Gaussian random numbers will be realized by taking the following probability measures,
\begin{equation}\label{eq:RectangularGGMeasure}
\dd \mu \left( \mathbf{A}_{l} \right) \propto \ee^{- \frac{\sqrt{N_{l} N_{l + 1}}}{\sigma_{l}^{2}} \Tr \left( \mathbf{A}_{l}^{\dagger} \mathbf{A}_{l} \right)} \DD \mathbf{A}_{l} .
\end{equation}
It is almost identical to (\ref{eq:SquareGGMeasure}), with the exception of the scaling of the variance, whose inverse has generically to be chosen of the same order of magnitude as the dimensions of the matrices in question --- in order to yield in the thermodynamic limit a meaningful density of the eigenvalues --- but otherwise it can be arbitrary; here, we decide for the factor of \smash{$\sqrt{N_{l} N_{l + 1}}$}. We will assume \smash{$N_{L + 1} = N_{1}$} (\ieNotAPPB \smash{$R_{1} = 1$}) when considering the eigenvalues of $\mathbf{P}$ (\ref{eq:MPBasicEquation}), but for the singular values (\ieNotAPPB the square roots of the eigenvalues of $\mathbf{Q}$ (\ref{eq:MQBasicEquation})), this requirement could be dropped.


\subsubsection{The Main Results of the Paper}
\label{sss:TheMainResultsOfThePaper}

The first main result of our calculation, which we accomplish in section~\ref{s:TheEigenvaluesOfAProductOfRectangularGaussianRandomMatrices} by means of non--Hermitian planar diagrammatics, is the following simple equation for the non--holomorphic $M$--transform of $\mathbf{P}$,
\begin{equation}\label{eq:MPBasicEquation}
\prod_{l = 1}^{L} \left( \frac{M_{\mathbf{P}} ( z , \overline{z} )}{R_{l}} + 1 \right) = \frac{| z |^{2}}{\sigma^{2}} .
\end{equation}
As already announced, the full information about the mean distribution of the eigenvalues of $\mathbf{P}$ can be retrieved in a straightforward way from this $M$--transform (\ref{eq:NonHolomorphicMTransformDefinition}), (\ref{eq:MeanSpectralDensityFromNonHolomorphicGreensFunction}). We see that (\ref{eq:MPBasicEquation}) is a polynomial equation of order $L$. Moreover, since the RHS is a function of \smash{$| z |^{2}$} only, so will the non--holomorphic $M$--transform be,
\begin{equation}\label{eq:MPAsAFunctionOfTheRadius}
M_{\mathbf{P}} ( z , \overline{z} ) = \mathfrak{M}_{\mathbf{P}} \left( | z |^{2} \right) .
\end{equation}
Consequently, the spectral density will display rotational symmetry, in concord with the square case (\ref{eq:SquareGGProductMeanSpectralDensity}). Finally, remark that the linear scale of the distribution is still given by $\sigma$ (\ref{eq:Sigma}), \ieNotAPPB $z$ appears only as the combination $z / \sigma$.

The second main contribution of the article, worked out in section~\ref{s:TheSingularValuesOfAProductOfRectangularGaussianRandomMatrices} with aid of the FRV multiplication law, is an equation obeyed by the (usual, Hermitian) $M$--transform of $\mathbf{Q}$ (\ref{eq:SingularValuesDefinition}), yielding thus (\ref{eq:MTransformDefinition}), (\ref{eq:MeanSpectralDensityFromGreensFunction}) the singular values squared of the product $\mathbf{P}$,
\begin{equation}\label{eq:MQBasicEquation}
\sqrt{R_{1}} \frac{M_{\mathbf{Q}} ( z ) + 1}{M_{\mathbf{Q}} ( z )} \prod_{l = 1}^{L} \left( \frac{M_{\mathbf{Q}} ( z )}{R_{l}} + 1 \right) = \frac{z}{\sigma^{2}} .
\end{equation}
This is a polynomial equation of order $( L + 1 )$. This equation for the product of square matrices has been derived in~\cite{BanicaBelinschiCapitaineCollins2007,BenaychGeorges2008}, while for rectangular ones, without our knowledge, in~\cite{Muller2002} in the context of wireless telecommunication; we thus present our derivation of the latter result as well.

As straightforward consequences of the main equations (\ref{eq:MPBasicEquation}) and (\ref{eq:MQBasicEquation}), we unravel the nature of the singularities of the mean spectral densities of $\mathbf{P}$ and $\mathbf{Q}$ at zero. We show that their behaviors at the origin are solely determined by the number $s$ of the rectangularity ratios (\ref{eq:ThermodynamicLimit}) equal to $1$, \ieNotAPPB
\begin{equation}\label{eq:s}
s \equiv \# \left\{ l = 1 , 2 , \ldots , L : N_{l} = N_{L + 1} \right\} = 1 , 2 , \ldots , L ,
\end{equation}
and are given by
\begin{equation}\label{eq:RhoPSingularityAtZero}
\rho_{\mathbf{P}} ( \lambda , \overline{\lambda} ) \sim | \lambda |^{- 2 \left( 1 - \frac{1}{s} \right)} , \quad \textrm{as} \quad \lambda \to 0 ,
\end{equation}
and
\begin{equation}\label{eq:RhoQSingularityAtZero}
\rho_{\mathbf{Q}} ( \lambda ) \sim \lambda^{- \frac{s}{s + 1}} , \quad \textrm{as} \quad \lambda \to 0 .
\end{equation}
Remark that when all the constituent matrices are square, $s = L$, then (\ref{eq:RhoPSingularityAtZero}) precisely reproduces the density in the entire domain of the spectrum (\ref{eq:SquareGGProductMeanSpectralDensity}), not only close to the origin.

Let us close with the following observation: (\ref{eq:MQBasicEquation}) happens to be remarkably similar to (\ref{eq:MPBasicEquation}): Setting \smash{$R_{1} = 1$}, which is a must if we wish to simultaneously compute both the eigenvalues (\ref{eq:MPBasicEquation}) and singular values (\ref{eq:MQBasicEquation}), one notices that it is sufficient to replace on the LHS \smash{$M_{\mathbf{P}} ( z , \overline{z} )$} by \smash{$M_{\mathbf{Q}} ( z )$}, and multiply by \smash{$( M_{\mathbf{Q}} ( z ) + 1 ) / M_{\mathbf{Q}} ( z )$}, while on the RHS to replace \smash{$| z |^{2}$} by $z$ --- in order to proceed from (\ref{eq:MPBasicEquation}) to (\ref{eq:MQBasicEquation}).

This unexpected striking relationship has led us to propose the following conjecture: If $\mathbf{X}$ is any non--Hermitian random matrix model whose mean spectrum possesses \emph{rotational symmetry}, \ieNotAPPB equivalently, whose non--holomorphic $M$--transform depends only on \smash{$| z |^{2}$} (\ref{eq:MPAsAFunctionOfTheRadius}), \smash{$M_{\mathbf{X}} ( z , \overline{z} ) = \mathfrak{M}_{\mathbf{X}} ( | z |^{2} )$} --- then, introducing the ``rotationally--symmetric non--holomorphic $N$--transform'' as the functional inverse
\begin{equation}\label{eq:RotationallySymmetricNonHolomorphicNTransformDefinition}
\mathfrak{M}_{\mathbf{X}} \left( \mathfrak{N}_{\mathbf{X}} ( z ) \right) = z ,
\end{equation}
its relation to the (usual) $N$--transform of the Hermitian matrix \smash{$\mathbf{X}^{\dagger} \mathbf{X}$} reads
\begin{equation}\label{eq:Conjecture}
N_{\mathbf{X}^{\dagger} \mathbf{X}} ( z ) = \frac{z + 1}{z} \mathfrak{N}_{\mathbf{X}} ( z ) .
\end{equation}
One may intuitively expect the existence of such a link, because a rotationally--symmetric spectral distribution is effectively one--dimensional, depending only on the absolute value squared of its complex argument --- and the eigenvalues of \smash{$\mathbf{X}^{\dagger} \mathbf{X}$} are precisely the absolute values squared of the eigenvalues of $\mathbf{X}$. {}From the practical point of view, typically, one of the two sides of (\ref{eq:Conjecture}) will be much easier to appropriate analytically, thus providing the other one for free. We postpone verification of this hypothesis for future work.


\subsubsection{The Plan of the Paper}
\label{sss:ThePlanOfThePaper}

The material in the article is divided into three levels of depth:
\begin{itemize}
\item Experts in the field may remain just with the above~\S\ref{sss:TheMainResultsOfThePaper}, which summarizes all of the results we have obtained in this work.
\item Intermediate--level readers may want moreover to consult sections~\ref{s:TheEigenvaluesOfAProductOfRectangularGaussianRandomMatrices} and~\ref{s:TheSingularValuesOfAProductOfRectangularGaussianRandomMatrices}, in which we compute the mean spectral densities of the product $\mathbf{P}$ (\ref{eq:ProductDefinition}) of rectangular Gaussian matrices (\ref{eq:RectangularGGMeasure}), as well as of \smash{$\mathbf{Q} \equiv \mathbf{P}^{\dagger} \mathbf{P}$} (\ref{eq:SingularValuesDefinition}), respectively.
\item Students are given furthermore appendices~\ref{s:TheDiagrammaticApproachToSolvingMatrixModels} and~\ref{s:FreeRandomVariablesInANutshell}, which introduce in a detailed and self--contained way non--Hermitian planar diagrammatics and free random variables calculus, respectively.
\end{itemize}
Section~\ref{s:Conclusions} concludes the paper and discusses some possible applications of its discoveries.


\section{The Eigenvalues of a Product of Rectangular Gaussian Random Matrices}
\label{s:TheEigenvaluesOfAProductOfRectangularGaussianRandomMatrices}


\subsection{Derivation}
\label{ss:EIGDerivation}

In this subsection, we present a derivation of the first main finding of this article, (\ref{eq:MPBasicEquation}), \ieNotAPPB an $L$--th--order polynomial equation for the non--holomorphic $M$--transform --- an object containing the full information about the mean spectrum of the (complex) eigenvalues (\ref{eq:NonHolomorphicMTransformDefinition}), (\ref{eq:MeanSpectralDensityFromNonHolomorphicGreensFunction}) --- of the (non--Hermitian) product $\mathbf{P}$ (\ref{eq:ProductDefinition}) of rectangular (\ref{eq:ThermodynamicLimit}) --- with the necessary constraint \smash{$N_{L + 1} = N_{1}$} --- Gaussian random matrices (\ref{eq:RectangularGGMeasure}). To this end, we employ an efficient technique of summing planar diagrams, called the ``Dyson--Schwinger's equations,'' extended to the non--Hermitian sector, and specifically suited to working with products of random matrices (a reader not familiar with the subject is strongly advised to consult appendix~\ref{s:TheDiagrammaticApproachToSolvingMatrixModels} for a didactic overview).


\subsubsection{The Linearization of the Product}
\label{sss:TheLinearizationOfTheProduct}

If one aims at computing the matrix--valued Green's function (\ref{eq:MatrixValuedGreensFunctionDefinition}) for a \emph{product} $\mathbf{P}$ (\ref{eq:ProductDefinition}) of random ensembles by means of planar diagrammatics, one notices that the problem becomes non--linear w.r.t. the constituent matrices. Opportunely, it is possible to linearize it by making use of the following trick: One trades the \smash{$N_{1} \times N_{1}$} matrix $\mathbf{P}$ for the \smash{$N_{\tot} \times N_{\tot}$} --- where \smash{$N_{\tot} \equiv N_{1} + N_{2} + \ldots + N_{L}$} --- matrix
\begin{equation}\label{eq:EIG01}
\widetilde{\mathbf{P}} \equiv \left( \begin{array}{ccccc} \Zero & \mathbf{A}_{1} & \Zero & \ldots & \Zero \\ \Zero & \Zero & \mathbf{A}_{2} & \ldots & \Zero \\ \vdots & \vdots & \vdots & \ddots & \vdots \\ \Zero & \Zero & \Zero & \ldots & \mathbf{A}_{L - 1} \\ \mathbf{A}_{L} & \Zero & \Zero & \ldots & \Zero \end{array} \right) ,
\end{equation}
with an $L \times L$ block structure, where the $\Zero$'s represent matrices of various appropriate dimensions entirely filled with zeros (we will also use the notation of \smash{$\Zero_{N}$} for the $N \times N$ zero matrix).

Now, the non--holomorphic $M$--transforms of $\mathbf{P}$ and \smash{$\widetilde{\mathbf{P}}$} are related in a simple way,
\begin{equation}\label{eq:EIG07a}
M_{\widetilde{\mathbf{P}}} ( w , \overline{w} ) = \frac{L N_{1}}{N_{\tot}} M_{\mathbf{P}} \left( w^{L} , \overline{w}^{L} \right) .
\end{equation}
We present the proof in \S\ref{sss:NonHermitianPlanarDiagrammaticsForProductsOfRandomMatrices}. In other words, solving the spectral problem for \smash{$\widetilde{\mathbf{P}}$} (\ref{eq:EIG01}) is equivalent to solving the original model $\mathbf{P}$ (\ref{eq:ProductDefinition}). The former is already linear w.r.t. the constituent matrices, and permits the Dyson--Schwinger's approach to calculating \smash{$M_{\widetilde{\mathbf{P}}} ( w , \overline{w} )$}, from which the desired \smash{$M_{\mathbf{P}} ( z , \overline{z} )$} is obtained with help of (\ref{eq:EIG07a}), albeit for a price of enlarging the size of the random matrix from \smash{$N_{1}$} to \smash{$N_{\tot}$}.


\subsubsection{The Dyson--Schwinger's Equations}
\label{sss:TheDysonSchwingersEquations}

As demonstrated in \S\ref{sss:TheNonHolomorphicGreensFunctionForTheGirkoGinibreEnsembleFromPlanarDiagrams}, the first step in writing the Dyson--Schwinger's equations is to know the propagators of the random matrix in question, namely, \smash{$\widetilde{\mathbf{P}}$}, or more precisely, its ``duplicated'' version (\ref{eq:MatrixValuedGreensFunctionMatrixDefinition1}),
\begin{equation}\label{eq:EIG09}
\widetilde{\mathbf{P}}^{\DD} = \left( \begin{array}{cc} \widetilde{\mathbf{P}} & \Zero \\ \Zero & \widetilde{\mathbf{P}}^{\dagger} \end{array} \right) .
\end{equation}
It is a \smash{$2 N_{\tot} \times 2 N_{\tot}$} matrix, but we will think of it as having four blocks (as distinguished in (\ref{eq:EIG09})), each being an $L \times L$ block matrix (see (\ref{eq:EIG01})). We will denote the $L \times L$ block indices in these four blocks by $l m$ (upper left corner), \smash{$l \overline{m}$} (upper right), \smash{$\overline{l} m$} (lower left), \smash{$\overline{l} \overline{m}$} (lower right), each one covering the range $1 , 2 , \ldots , L$ (compare (\ref{eq:GGDPropagators})); for example, \smash{$[ \widetilde{\mathbf{P}}^{\DD} ]_{\overline{2} \overline{1}} = \mathbf{A}_{1}^{\dagger}$}. All the other featured matrices will inherit this structure, for instance,
$$
\mathbf{G}^{\DD} = \left( \begin{array}{c|c} \mathbf{G}^{w w} & \mathbf{G}^{w \overline{w}} \\ \hline \mathbf{G}^{\overline{w} w} & \mathbf{G}^{\overline{w} \overline{w}} \end{array} \right) =
$$
\begin{equation}\label{eq:EIG10}
= \left( \begin{array}{cccc|cccc} \left[ \mathbf{G}^{\DD} \right]_{1 1} & \left[ \mathbf{G}^{\DD} \right]_{1 2} & \ldots & \left[ \mathbf{G}^{\DD} \right]_{1 L} & \left[ \mathbf{G}^{\DD} \right]_{1 \overline{1}} & \left[ \mathbf{G}^{\DD} \right]_{1 \overline{2}} & \ldots & \left[ \mathbf{G}^{\DD} \right]_{1 \overline{L}} \\ \left[ \mathbf{G}^{\DD} \right]_{2 1} & \left[ \mathbf{G}^{\DD} \right]_{2 2} & \ldots & \left[ \mathbf{G}^{\DD} \right]_{2 L} & \left[ \mathbf{G}^{\DD} \right]_{2 \overline{1}} & \left[ \mathbf{G}^{\DD} \right]_{2 \overline{2}} & \ldots & \left[ \mathbf{G}^{\DD} \right]_{2 \overline{L}} \\ \vdots & \vdots & \ddots & \vdots & \vdots & \vdots & \ddots & \vdots \\ \left[ \mathbf{G}^{\DD} \right]_{L 1} & \left[ \mathbf{G}^{\DD} \right]_{L 2} & \ldots & \left[ \mathbf{G}^{\DD} \right]_{L L} & \left[ \mathbf{G}^{\DD} \right]_{L \overline{1}} & \left[ \mathbf{G}^{\DD} \right]_{L \overline{2}} & \ldots & \left[ \mathbf{G}^{\DD} \right]_{L \overline{L}} \\ \hline \left[ \mathbf{G}^{\DD} \right]_{\overline{1} 1} & \left[ \mathbf{G}^{\DD} \right]_{\overline{1} 2} & \ldots & \left[ \mathbf{G}^{\DD} \right]_{\overline{1} L} & \left[ \mathbf{G}^{\DD} \right]_{\overline{1} \overline{1}} & \left[ \mathbf{G}^{\DD} \right]_{\overline{1} \overline{2}} & \ldots & \left[ \mathbf{G}^{\DD} \right]_{\overline{1} \overline{L}} \\ \left[ \mathbf{G}^{\DD} \right]_{\overline{2} 1} & \left[ \mathbf{G}^{\DD} \right]_{\overline{2} 2} & \ldots & \left[ \mathbf{G}^{\DD} \right]_{\overline{2} L} & \left[ \mathbf{G}^{\DD} \right]_{\overline{2} \overline{1}} & \left[ \mathbf{G}^{\DD} \right]_{\overline{2} \overline{2}} & \ldots & \left[ \mathbf{G}^{\DD} \right]_{\overline{2} \overline{L}} \\ \vdots & \vdots & \ddots & \vdots & \vdots & \vdots & \ddots & \vdots \\ \left[ \mathbf{G}^{\DD} \right]_{\overline{L} 1} & \left[ \mathbf{G}^{\DD} \right]_{\overline{L} 2} & \ldots & \left[ \mathbf{G}^{\DD} \right]_{\overline{L} L} & \left[ \mathbf{G}^{\DD} \right]_{\overline{L} \overline{1}} & \left[ \mathbf{G}^{\DD} \right]_{\overline{L} \overline{2}} & \ldots & \left[ \mathbf{G}^{\DD} \right]_{\overline{L} \overline{L}} \end{array} \right) ,
\end{equation}
and similarly for \smash{$\mathbf{W}^{\DD}$} and \smash{$\mathbf{\Sigma}^{\DD}$}. (We use $w$ instead of $z$ to comply with the notation in (\ref{eq:EIG07a}), as well as disregard for simplicity all the subscripts and the symbols of dependence on $w , \overline{w}$.)

We are interested in computing the non--holomorphic Green's function of \smash{$\widetilde{\mathbf{P}}$} (\ref{eq:MatrixValuedGreensFunctionBlockZZ}), \ieNotAPPB
\begin{equation}\label{eq:EIG11}
G_{\widetilde{\mathbf{P}}} ( w , \overline{w} ) = \frac{1}{N_{\tot}} \Tr \mathbf{G}^{w w} = \frac{1}{N_{\tot}} \sum_{l = 1}^{L} \Tr \left[ \mathbf{G}^{\DD} \right]_{l l} = \frac{1}{N_{\tot}} \sum_{l = 1}^{L} N_{l} \mathcal{G}_{l l} ,
\end{equation}
where it is useful to define the normalized traces
$$
\mathcal{G}_{l l} \equiv \frac{1}{N_{l}} \Tr \left[ \mathbf{G}^{\DD} \right]_{l l} , \quad \mathcal{G}_{l \overline{l}} \equiv \frac{1}{N_{l}} \Tr \left[ \mathbf{G}^{\DD} \right]_{l \overline{l}} ,
$$
\begin{equation}\label{eq:EIG12}
\mathcal{G}_{\overline{l} l} \equiv \frac{1}{N_{l}} \Tr \left[ \mathbf{G}^{\DD} \right]_{\overline{l} l} , \quad \mathcal{G}_{\overline{l} \overline{l}} \equiv \frac{1}{N_{l}} \Tr \left[ \mathbf{G}^{\DD} \right]_{\overline{l} \overline{l}} .
\end{equation}
Hence, we should find the \smash{$\mathcal{G}_{l l}$}'s.

The $2$--point correlation functions of the ensembles \smash{$\mathbf{A}_{l}$} are readily visible from their probability measures (\ref{eq:RectangularGGMeasure}),
\begin{equation}\label{eq:EIG13}
\la \left[ \mathbf{A}_{l} \right]_{a b} \left[ \mathbf{A}_{m}^{\dagger} \right]_{c d} \ra = \frac{\sigma_{l}^{2}}{\sqrt{N_{l} N_{l + 1}}} \delta_{l m} \delta_{a d} \delta_{b c} ,
\end{equation}
with all the rest equal to zero. In terms of \smash{$\widetilde{\mathbf{P}}^{\DD}$}, (\ref{eq:EIG13}) means that the only non--zero propagators are
\begin{align}
\la \left[ \widetilde{\mathbf{P}}^{\DD} \right]_{1 2} \left[ \widetilde{\mathbf{P}}^{\DD} \right]_{\overline{2} \overline{1}} \ra & = \frac{\sigma_{1}^{2}}{\sqrt{N_{1} N_{2}}} \Id_{N_{1}} \otimes \Id_{N_{2}} , \nonumber\\
\la \left[ \widetilde{\mathbf{P}}^{\DD} \right]_{2 3} \left[ \widetilde{\mathbf{P}}^{\DD} \right]_{\overline{3} \overline{2}} \ra & = \frac{\sigma_{2}^{2}}{\sqrt{N_{2} N_{3}}} \Id_{N_{2}} \otimes \Id_{N_{3}} , \nonumber\\
& \vdots \nonumber\\
\la \left[ \widetilde{\mathbf{P}}^{\DD} \right]_{L 1} \left[ \widetilde{\mathbf{P}}^{\DD} \right]_{\overline{1} \overline{L}} \ra & = \frac{\sigma_{L}^{2}}{\sqrt{N_{L} N_{1}}} \Id_{N_{L}} \otimes \Id_{N_{1}} , \label{eq:EIG14}
\end{align}
where the tensor notation is a shortcut for what could be alternatively achieved using a double--index notation,
\begin{equation}\label{eq:EIG15}
\la \left[ \widetilde{\mathbf{P}}^{\DD} \right]_{( l , a ) , ( l + 1 , A )} \left[ \widetilde{\mathbf{P}}^{\DD} \right]_{( \overline{l + 1} , B ) , ( \overline{l} , b )} \ra = \frac{\sigma_{l}^{2}}{\sqrt{N_{l} N_{l + 1}}} \delta_{a b} \delta_{A B} ,
\end{equation}
where $l = 1 , 2 , \ldots , L$ (and we use a cyclic identification convention here, $L + 1 = 1$), \smash{$a , b = 1 , 2 , \ldots , N_{l}$}, and \smash{$A , B = 1 , 2 , \ldots , N_{l + 1}$}.

Thus, we are now in position to write down the two Dyson--Schwinger's equations for \smash{$\widetilde{\mathbf{P}}^{\DD}$}. The first one, being in fact the definition of the self--energy matrix, is independent of the propagators and identical to (\ref{eq:GGDysonSchwinger1}) and (\ref{eq:GUEDysonSchwinger1}),
\begin{equation}\label{eq:EIG16}
\mathbf{G}^{\DD} = \left( \mathbf{W}^{\DD} - \mathbf{\Sigma}^{\DD} \right)^{- 1} .
\end{equation}
The second one is pictorially presented above formula (\ref{eq:GUEDysonSchwinger2}), and the structure of the propagators (\ref{eq:EIG14}) implies that the only non--zero blocks of the self--energy matrix read
\begin{equation}\label{eq:EIG17}
\left[ \mathbf{\Sigma}^{\DD} \right]_{l \overline{l}} = \frac{\sigma_{l}^{2}}{\sqrt{N_{l} N_{l + 1}}} \Tr \left[ \mathbf{G}^{\DD} \right]_{l + 1 , \overline{l + 1}} \Id_{N_{l}} = \underbrace{\sigma_{l}^{2} \sqrt{\frac{N_{l + 1}}{N_{l}}} \mathcal{G}_{l + 1 , \overline{l + 1}}}_{\equiv \alpha_{l}} \Id_{N_{l}} ,
\end{equation}
\begin{equation}\label{eq:EIG18}
\left[ \mathbf{\Sigma}^{\DD} \right]_{\overline{l} l} = \frac{\sigma_{l - 1}^{2}}{\sqrt{N_{l - 1} N_{l}}} \Tr \left[ \mathbf{G}^{\DD} \right]_{\overline{l - 1} , l - 1} \Id_{N_{l}} = \underbrace{\sigma_{l - 1}^{2} \sqrt{\frac{N_{l - 1}}{N_{l}}} \mathcal{G}_{\overline{l - 1} , l - 1}}_{\equiv \beta_{l}} \Id_{N_{l}} ,
\end{equation}
for all $l = 1 , 2 , \ldots , L$, with the cyclic convention $0 = L$, where the normalized traces (\ref{eq:EIG12}) have been used.

Results (\ref{eq:EIG17}), (\ref{eq:EIG18}) mean that the four blocks of the matrix \smash{$( \mathbf{W}^{\DD} - \mathbf{\Sigma}^{\DD} )$} are diagonal,
$$
\mathbf{W}^{\DD} - \mathbf{\Sigma}^{\DD} =
$$
\begin{equation}\label{eq:EIG19}
= \left( \begin{array}{cccc|cccc} w \Id_{N_{1}} & \Zero & \ldots & \Zero & - \alpha_{1} \Id_{N_{1}} & \Zero & \ldots & \Zero \\ \Zero & w \Id_{N_{2}} & \ldots & \Zero & \Zero & - \alpha_{2} \Id_{N_{2}} & \ldots & \Zero \\ \vdots & \vdots & \ddots & \vdots & \vdots & \vdots & \ddots & \vdots \\ \Zero & \Zero & \ldots & w \Id_{N_{L}} & \Zero & \Zero & \ldots & - \alpha_{L} \Id_{N_{L}} \\ \hline - \beta_{1} \Id_{N_{1}} & \Zero & \ldots & \Zero & \overline{w} \Id_{N_{1}} & \Zero & \ldots & \Zero \\ \Zero & - \beta_{2} \Id_{N_{2}} & \ldots & \Zero & \Zero & \overline{w} \Id_{N_{2}} & \ldots & \Zero \\ \vdots & \vdots & \ddots & \vdots & \vdots & \vdots & \ddots & \vdots \\ \Zero & \Zero & \ldots & - \beta_{L} \Id_{N_{L}} & \Zero & \Zero & \ldots & \overline{w} \Id_{N_{L}} \end{array} \right) .
\end{equation}
Such a matrix can be straightforwardly inverted: its four blocks remain diagonal, and read
$$
\left( \mathbf{W}^{\DD} - \mathbf{\Sigma}^{\DD} \right)^{- 1} =
$$
\begin{equation}\label{eq:EIG20}
= \left( \begin{array}{cccc|cccc} \overline{w} \gamma_{1} \Id_{N_{1}} & \Zero & \ldots & \Zero & \alpha_{1} \gamma_{1} \Id_{N_{1}} & \Zero & \ldots & \Zero \\ \Zero & \overline{w} \gamma_{2} \Id_{N_{2}} & \ldots & \Zero & \Zero & \alpha_{2} \gamma_{2} \Id_{N_{2}} & \ldots & \Zero \\ \vdots & \vdots & \ddots & \vdots & \vdots & \vdots & \ddots & \vdots \\ \Zero & \Zero & \ldots & \overline{w} \gamma_{L} \Id_{N_{L}} & \Zero & \Zero & \ldots & \alpha_{L} \gamma_{L} \Id_{N_{L}} \\ \hline \beta_{1} \gamma_{1} \Id_{N_{1}} & \Zero & \ldots & \Zero & w \gamma_{1} \Id_{N_{1}} & \Zero & \ldots & \Zero \\ \Zero & \beta_{2} \gamma_{2} \Id_{N_{2}} & \ldots & \Zero & \Zero & w \gamma_{2} \Id_{N_{2}} & \ldots & \Zero \\ \vdots & \vdots & \ddots & \vdots & \vdots & \vdots & \ddots & \vdots \\ \Zero & \Zero & \ldots & \beta_{L} \gamma_{L} \Id_{N_{L}} & \Zero & \Zero & \ldots & w \gamma_{L} \Id_{N_{L}} \end{array} \right) ,
\end{equation}
where for short, for all $l = 1 , 2 , \ldots , L$,
\begin{equation}\label{eq:EIG21}
\frac{1}{\gamma_{l}} \equiv | w |^{2} - \alpha_{l} \beta_{l} = | w |^{2} - \left( \sigma_{l - 1} \sigma_{l} \right)^{2} \frac{\sqrt{N_{l - 1} N_{l + 1}}}{N_{l}} \mathcal{G}_{l + 1 , \overline{l + 1}} \mathcal{G}_{\overline{l - 1} , l - 1} .
\end{equation}
Substituting (\ref{eq:EIG20}), which is an implication of the second Dyson--Schwinger's equation, to the first one (\ref{eq:EIG16}), we discover that the only non--zero blocks of the duplicated Green's function (\ref{eq:EIG10}) are, for all $l = 1 , 2 , \ldots , L$,
$$
\left[ \mathbf{G}^{\DD} \right]_{l l} = \overline{w} \gamma_{l} \Id_{N_{l}} , \quad \left[ \mathbf{G}^{\DD} \right]_{l \overline{l}} = \alpha_{l} \gamma_{l} \Id_{N_{l}} ,
$$
\begin{equation}\label{eq:EIG22}
\left[ \mathbf{G}^{\DD} \right]_{\overline{l} l} = \beta_{l} \gamma_{l} \Id_{N_{l}} , \quad \left[ \mathbf{G}^{\DD} \right]_{\overline{l} \overline{l}} = w \gamma_{l} \Id_{N_{l}} .
\end{equation}
Taking the normalized traces of both sides of every equality in (\ref{eq:EIG22}), this leads to the final set of equations,
\begin{equation}\label{eq:EIG23}
\mathcal{G}_{l l} = \overline{w} \gamma_{l} , \quad \mathcal{G}_{l \overline{l}} = \alpha_{l} \gamma_{l} , \quad \mathcal{G}_{\overline{l} l} = \beta_{l} \gamma_{l} , \quad \mathcal{G}_{\overline{l} \overline{l}} = w \gamma_{l} .
\end{equation}

To summarize, the structure of equations (\ref{eq:EIG23}) is the following: The fourth one, as a general caveat, is the conjugate of the first one, so it is redundant. The second and third ones read
\begin{equation}\label{eq:EIG24}
\mathcal{G}_{l \overline{l}} = \sigma_{l}^{2} \sqrt{\frac{N_{l + 1}}{N_{l}}} \mathcal{G}_{l + 1 , \overline{l + 1}} \gamma_{l} ,
\end{equation}
\begin{equation}\label{eq:EIG25}
\mathcal{G}_{\overline{l} l} = \sigma_{l - 1}^{2} \sqrt{\frac{N_{l - 1}}{N_{l}}} \mathcal{G}_{\overline{l - 1} , l - 1} \gamma_{l} .
\end{equation}
We see that (\ref{eq:EIG21}), (\ref{eq:EIG24}) and (\ref{eq:EIG25}) form a closed set of $3 L$ equations for $3 L$ unknowns, \smash{$\mathcal{G}_{l \overline{l}}$}, \smash{$\mathcal{G}_{\overline{l} l}$} and \smash{$\gamma_{l}$} --- these we will now (\S\ref{sss:SolvingTheDysonSchwingersEquations}) attempt to unfold. Once solved, in particular, when the \smash{$\gamma_{l}$}'s are found, the first expression in (\ref{eq:EIG23}) yields \smash{$\mathcal{G}_{l l}$}, \ieNotAPPB as a consequence (\ref{eq:EIG11}), the non--holomorphic Green's function of \smash{$\widetilde{\mathbf{P}}$}, and subsequently, the non--holomorphic $M$--transforms of \smash{$\widetilde{\mathbf{P}}$} (\ref{eq:NonHolomorphicMTransformDefinition}), as well as of $\mathbf{P}$ (in the argument \smash{$w^{L}$}) (\ref{eq:EIG07a}),
$$
G_{\widetilde{\mathbf{P}}} ( w , \overline{w} ) = \overline{w} \frac{1}{N_{\tot}} \sum_{l = 1}^{L} N_{l} \gamma_{l} , \quad \textrm{\ieNotAPPB} \quad M_{\widetilde{\mathbf{P}}} ( w , \overline{w} ) = \frac{1}{N_{\tot}} \sum_{l = 1}^{L} N_{l} \mu_{l} ,
$$
\begin{equation}\label{eq:EIG26}
\textrm{\ieNotAPPB} \quad M_{\mathbf{P}} \left( w^{L} , \overline{w}^{L} \right) = \frac{1}{L} \sum_{l = 1}^{L} R_{l} \mu_{l} ,
\end{equation}
where we have traded the \smash{$\gamma_{l}$}'s for a more convenient set of variables,
\begin{equation}\label{eq:EIG27}
\mu_{l} \equiv | w |^{2} \gamma_{l} - 1 ,
\end{equation}
and recall that the rectangularity ratios \smash{$R_{l} = N_{l} / N_{1}$} (\ref{eq:ThermodynamicLimit}).


\subsubsection{Solving the Dyson--Schwinger's Equations}
\label{sss:SolvingTheDysonSchwingersEquations}

Let us start from an observation that if we knew the \smash{$\gamma_{l}$}'s, then equations (\ref{eq:EIG24}) and (\ref{eq:EIG25}) would comprise a set of decoupled recurrence relations for \smash{$\mathcal{G}_{l \overline{l}}$} and \smash{$\mathcal{G}_{\overline{l} l}$}, respectively. Assume this is the case, and iterate these recurrences down to $l = 1$,
\begin{equation}\label{eq:EIG28}
\mathcal{G}_{l \overline{l}} = \mathcal{G}_{1 \overline{1}} \frac{1}{\left( \sigma_{1} \sigma_{2} \ldots \sigma_{l - 1} \right)^{2}} \frac{1}{\sqrt{R_{l}}} \frac{1}{\gamma_{1} \gamma_{2} \ldots \gamma_{l - 1}} ,
\end{equation}
\begin{equation}\label{eq:EIG29}
\mathcal{G}_{\overline{l} l} = \mathcal{G}_{\overline{1} 1} \left( \sigma_{1} \sigma_{2} \ldots \sigma_{l - 1} \right)^{2} \frac{1}{\sqrt{R_{l}}} \gamma_{2} \ldots \gamma_{l} .
\end{equation}

Actually, (\ref{eq:EIG28}) and (\ref{eq:EIG29}) hold true for all $l = 1 , 2 , \ldots , L$ (and not only $l \geq 2$), where the $l = 1$ case is obtained by applying the cyclic convention $0 = L$. This cyclic boundary constraint for (\ref{eq:EIG28}) is
\begin{equation}\label{eq:EIG30}
\left( \sigma_{1} \sigma_{2} \ldots \sigma_{L} \right)^{2} \left( \gamma_{1} \gamma_{2} \ldots \gamma_{L} \right) \mathcal{G}_{1 \overline{1}} = \mathcal{G}_{1 \overline{1}}
\end{equation}
(and analogously for (\ref{eq:EIG29}), with \smash{$\mathcal{G}_{1 \overline{1}}$} replaced by \smash{$\mathcal{G}_{\overline{1} 1}$}). One possibility now is that \smash{$\mathcal{G}_{1 \overline{1}} = 0$}. This implies in turn (\ref{eq:EIG28}) that for all $l$, \smash{$\mathcal{G}_{l \overline{l}} = 0$}, \ieNotAPPB furthermore (\ref{eq:EIG21}), \smash{$\gamma_{l} = 1 / | w |^{2}$}, or (\ref{eq:EIG27}), \smash{$\mu_{l} = 0$}, and therefore (\ref{eq:EIG26}), \smash{$M_{\mathbf{P}} ( z , \overline{z} ) = 0$}. This is thus the trivial holomorphic solution, valid outside of the mean eigenvalues' domain. We are however interested in the eigenvalues, \ieNotAPPB in the non--holomorphic solution, so let us take \smash{$\mathcal{G}_{1 \overline{1}} \neq 0$} henceforth. The single equation (\ref{eq:EIG30}) reduces then to
\begin{equation}\label{eq:EIG31}
\gamma_{1} \gamma_{2} \ldots \gamma_{L} = \frac{1}{\sigma^{2}} , \quad \textrm{\ieNotAPPB} \quad \left( \mu_{1} + 1 \right) \left( \mu_{2} + 1 \right) \ldots \left( \mu_{L} + 1 \right) = \frac{\left| w^{L} \right|^{2}}{\sigma^{2}} ,
\end{equation}
with $\sigma$ defined in (\ref{eq:Sigma}).

Substitute now (\ref{eq:EIG28}) and (\ref{eq:EIG29}) into (\ref{eq:EIG21}), and go to the variables \smash{$\mu_{l}$} --- after some simplifications, (\ref{eq:EIG21}) acquires the form, for all $l = 1 , 2 , \ldots , L$,
\begin{equation}\label{eq:EIG32}
\mu_{l} = \frac{| w |^{2} \mathcal{G}_{1 \overline{1}} \mathcal{G}_{\overline{1} 1}}{\mu_{1} + 1} \frac{1}{R_{l}} ,
\end{equation}
Take it first for $l = 1$,
\begin{equation}\label{eq:EIG33}
\mu_{1} = \frac{| w |^{2} \mathcal{G}_{1 \overline{1}} \mathcal{G}_{\overline{1} 1}}{\mu_{1} + 1} ,
\end{equation}
where we have used \smash{$R_{1} = 1$}, and replace the RHS of (\ref{eq:EIG33}) appearing in (\ref{eq:EIG32}) by its LHS --- obtaining thereby all the \smash{$\mu_{l}$}'s, for $l = 2 , 3 , \ldots , L$, as simply related to \smash{$\mu_{1}$},
\begin{equation}\label{eq:EIG34}
\mu_{l} = \frac{\mu_{1}}{R_{l}} .
\end{equation}
Hence, if we manage to compute \smash{$\mu_{1}$}, then all the \smash{$\mu_{l}$}'s will be known as well; this is done by plugging (\ref{eq:EIG34}) into (\ref{eq:EIG31}), which then becomes a polynomial equation of order $L$ for \smash{$\mu_{1}$},
\begin{equation}\label{eq:EIG35}
\left( \mu_{1} + 1 \right) \left( \frac{\mu_{1}}{R_{2}} + 1 \right) \ldots \left( \frac{\mu_{1}}{R_{L}} + 1 \right) = \frac{\left| w^{L} \right|^{2}}{\sigma^{2}} .
\end{equation}
This completes the solution of the fundamental set of equations (\ref{eq:EIG21}), (\ref{eq:EIG24}), (\ref{eq:EIG25}).

As mentioned, the knowledge of all the \smash{$\mu_{l}$}'s yields the desired non--holomorphic $M$--transform of $\mathbf{P}$; indeed, substituting (\ref{eq:EIG34}) into (\ref{eq:EIG26}), we simply get
\begin{equation}\label{eq:EIG36}
M_{\mathbf{P}} \left( w^{L} , \overline{w}^{L} \right) = \mu_{1} .
\end{equation}
In other words, it remains to change the complex argument from $w$ to \smash{$z = w^{L}$} in order to see that \smash{$M_{\mathbf{P}} ( z , \overline{z} )$} obeys the $L$--th--order polynomial equation
\begin{equation}\label{eq:EIG37}
\left( \frac{M_{\mathbf{P}} ( z , \overline{z} )}{R_{1}} + 1 \right) \left( \frac{M_{\mathbf{P}} ( z , \overline{z} )}{R_{2}} + 1 \right) \ldots \left( \frac{M_{\mathbf{P}} ( z , \overline{z} )}{R_{L}} + 1 \right) = \frac{| z |^{2}}{\sigma^{2}} ,
\end{equation}
which is precisely the first main result of our article, (\ref{eq:MPBasicEquation}).

The last point is to determine the domain of validity of the non--holomorphic solution (\ref{eq:EIG37}), \ieNotAPPB the domain where on average the eigenvalues of $\mathbf{P}$ lie. We know (\ref{eq:NonHolomorphicGreensFunctionOnTheBoundaryOfD}) that on the boundary of this domain, the non--holomorphic and holomorphic solutions must agree; plugging thus the latter (\smash{$M_{\mathbf{P}} ( z , \overline{z} ) = 0$}) into (\ref{eq:EIG37}), we obtain the equation of this borderline,
\begin{equation}\label{eq:EIG38}
| z | = \sigma .
\end{equation}
It means that the eigenvalues of $\mathbf{P}$ are scattered on average within a centered circle of radius $\sigma$, with the density stemming from (\ref{eq:EIG37}) by virtue of (\ref{eq:NonHolomorphicMTransformDefinition}), (\ref{eq:MeanSpectralDensityFromNonHolomorphicGreensFunction}).

\begin{figure}[ht]
\begin{center}
\includegraphics[width=8.5cm]{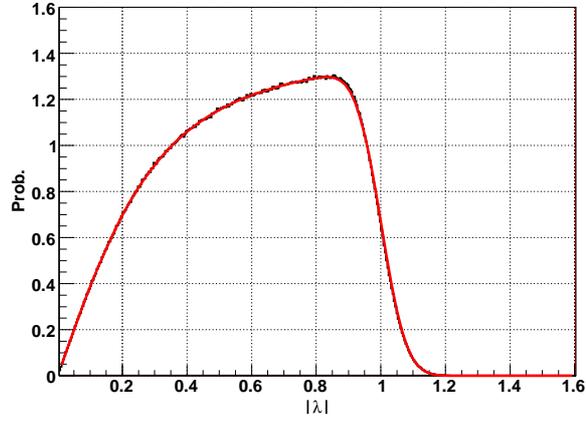}
\caption{Numerical verification of the theoretical formula (\ref{eq:EIG41}) for (the radial part (\ref{eq:EIG46}) of) the mean spectral density \smash{$\rho_{\mathbf{P}} ( z , \overline{z} )$} of the product $\mathbf{P}$ of $L = 2$ rectangular Gaussian random matrices, as well as the finite--size correction (\ref{eq:EIG47}). A numerical histogram (the black line) versus the theoretical prediction (\ref{eq:EIG41}), supplemented with the finite--size smoothing (\ref{eq:EIG47}) (the red plot), for \smash{$N_{1} = 100$} and \smash{$N_{2} = 200$} (\ieNotAPPB \smash{$R = R_{2} = 2$}), and for \smash{$10^{5}$} Monte--Carlo iterations (\ieNotAPPB the histogram is made of \smash{$10^{7}$}) eigenvalues. The adjustable parameter $q$ (\ref{eq:EIG47}) is fitted to be $q \approx 1.14$.}
\label{fig:EIGL2a}
\end{center}
\end{figure}

\begin{figure}[ht]
\begin{center}
\includegraphics[width=8.5cm]{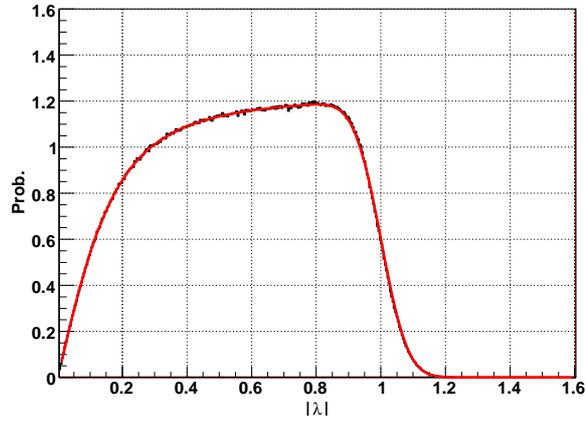}
\caption{An analogous graph to figure~\ref{fig:EIGL2a}, this time with \smash{$N_{1} = 100$} and \smash{$N_{2} = 150$} (\ieNotAPPB $R = 1.5$). We find $q \approx 1.08$ here.}
\label{fig:EIGL2b}
\end{center}
\end{figure}

\begin{figure}[ht]
\begin{center}
\includegraphics[width=8.5cm]{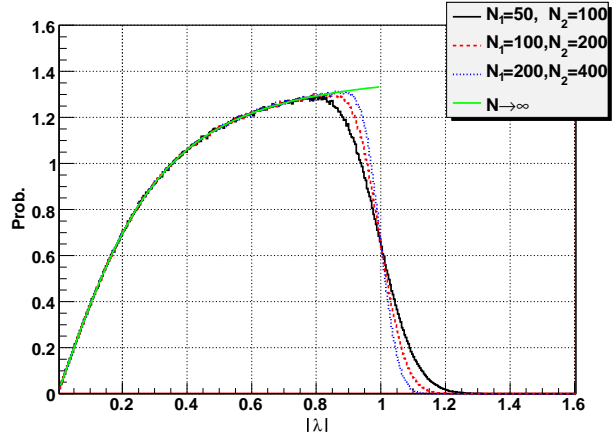}
\caption{An analysis of the finite--size effects: Numerical histograms for \smash{$N_{1} = 50$}, \smash{$N_{2} = 100$} (black), \smash{$N_{1} = 100$}, \smash{$N_{2} = 200$} (dashed red), \smash{$N_{1} = 200$}, \smash{$N_{2} = 400$} (dotted blue), \ieNotAPPB with the same rectangularity ratio $R = 2$, but increasing matrix dimensions. We observe how these plots approach the green line of the theoretical formula (\ref{eq:EIG41}) for the density in the thermodynamic limit.}
\label{fig:EIGL2c}
\end{center}
\end{figure}

\begin{figure}[ht]
\begin{center}
\includegraphics[width=8.5cm]{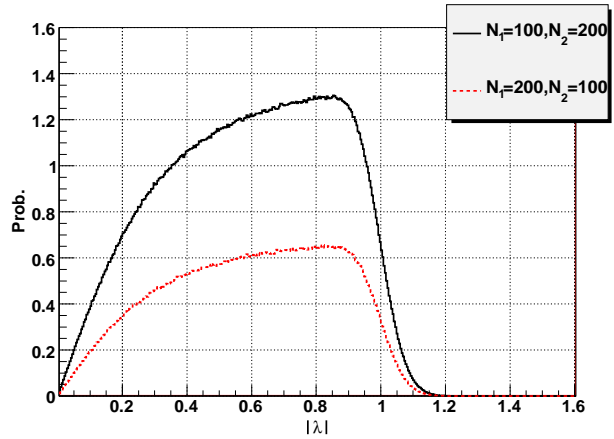}
\caption{Numerical histograms for the matrix sizes of \smash{$N_{1} = 100$}, \smash{$N_{2} = 200$} (\ieNotAPPB $R = 2$; black) and \smash{$N_{1} = 200$}, \smash{$N_{2} = 100$} (\ieNotAPPB $R = 1 / 2$; red). Due to the presence of the zero modes (not displayed in the picture), the latter is half of the former.}
\label{fig:EIGL2d}
\end{center}
\end{figure}

\begin{figure}[ht]
\begin{center}
\includegraphics[width=8.5cm]{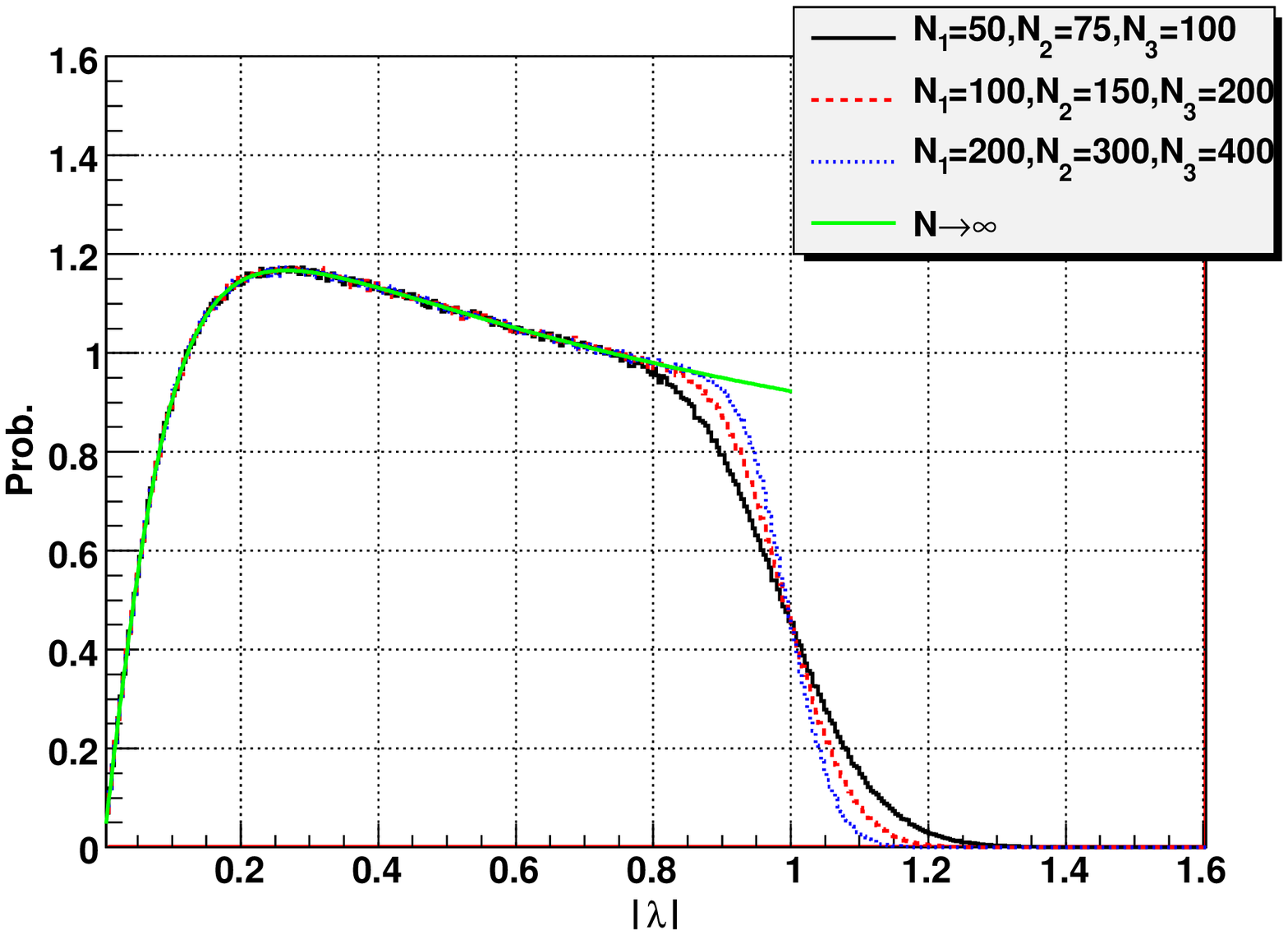}
\includegraphics[width=8.5cm]{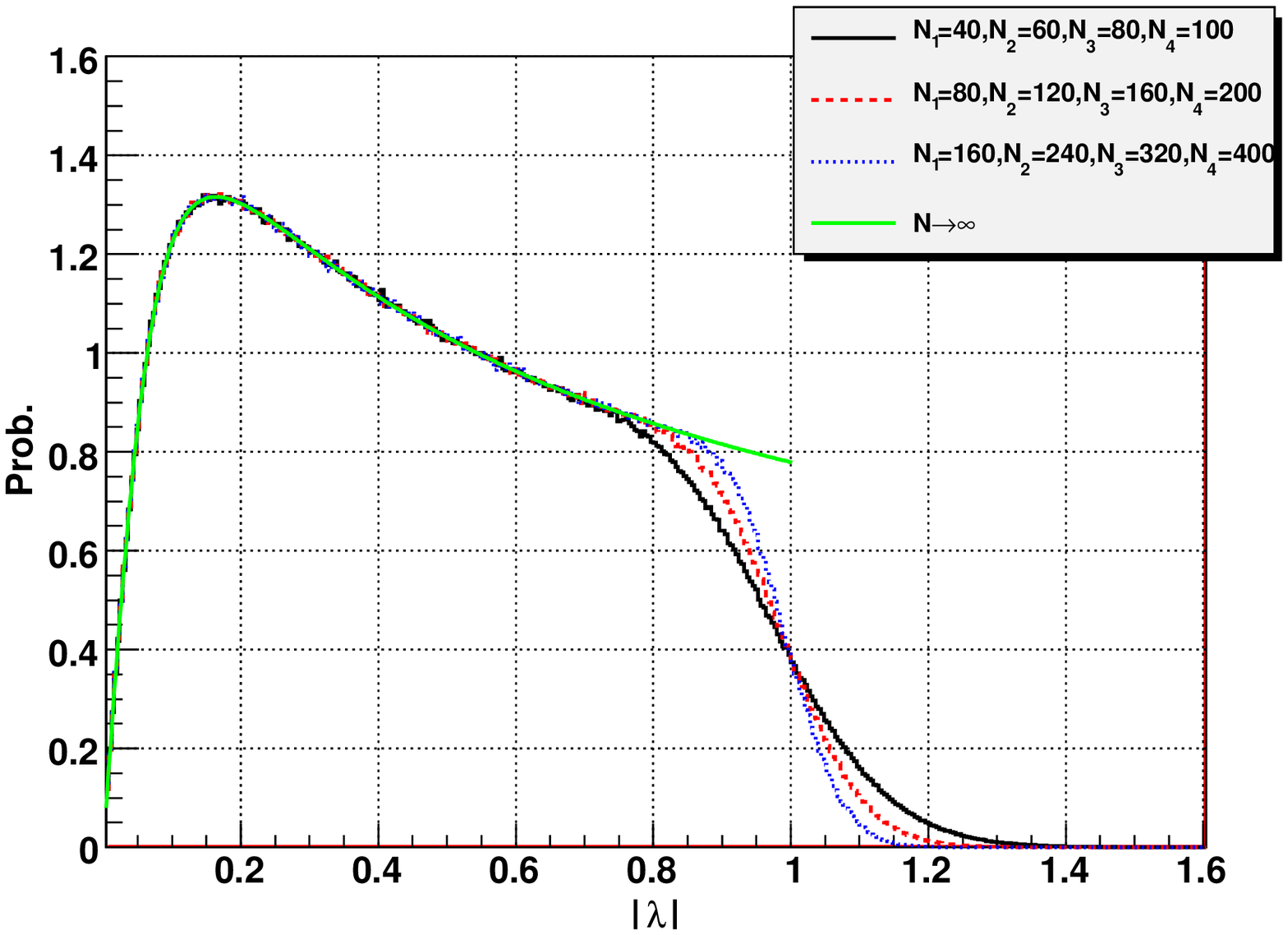}
\caption{Analogous graphs to figure~\ref{fig:EIGL2c}, but for $L = 3$ (up) and $L = 4$ (down).}
\label{fig:EIGL34}
\end{center}
\end{figure}


\subsection{Comments}
\label{ss:EIGComments}


\subsubsection{Example: $L = 2$}
\label{sss:EIGExampleLTwo}

Since our main equation (\ref{eq:EIG37}) is polynomial of order $L$, in the special case of $L = 2$ it will be easily solvable. Indeed, the non--holomorphic $M$--transform reads
\begin{equation}\label{eq:EIG39}
M_{\mathbf{P}} ( z , \overline{z} ) = \frac{1}{2} \left( - 1 - R + \sqrt{( 1 - R )^{2} + 4 R \frac{| z |^{2}}{\sigma^{2}}} \right) ,
\end{equation}
where we call \smash{$R \equiv R_{2} = N_{2} / N_{1}$} (the only non--trivial rectangularity ratio here), and where we have picked one solution out of the two roots of the quadratic equation (\ref{eq:EIG37}) in such a way that it satisfies \smash{$M_{\mathbf{P}} ( z , \overline{z} ) |_{| z | = \sigma} = 0$}, \ieNotAPPB the matching condition with the holomorphic solution on the borderline (\ref{eq:EIG38}). As a result, we immediately arrive at the non--holomorphic Green's function (\ref{eq:NonHolomorphicMTransformDefinition}),
\begin{equation}\label{eq:EIG40}
G_{\mathbf{P}} ( z , \overline{z} ) = \frac{1}{2 z} \left( 1 - R + \sqrt{( 1 - R )^{2} + 4 R \frac{| z |^{2}}{\sigma^{2}}} \right) .
\end{equation}

The average density of the eigenvalues stems from (\ref{eq:EIG40}) by taking the derivative w.r.t. \smash{$\overline{z}$} (\ref{eq:MeanSpectralDensityFromNonHolomorphicGreensFunction}), however, one has to be cautious in the vicinity of zero in order to properly take into account the zero modes. Let us first expand (\ref{eq:EIG40}) near $z = 0$ for the purpose of making its behavior clearly visible,
\begin{equation}\label{eq:EIG40a}
G_{\mathbf{P}} ( z , \overline{z} ) \sim \frac{f}{z} + \textrm{regular terms} , \quad \textrm{where} \quad f \equiv \left\{ \begin{array}{ll} 1 - R , & \quad \textrm{for} \quad R < 1 , \\ 0 , & \quad \textrm{for} \quad R \geq 1 . \end{array} \right.
\end{equation}
Applying the derivative \smash{$( 1 / \pi ) \partial_{\overline{z}}$} to this singular term produces the complex Dirac's delta at the origin, \smash{$f \delta^{( 2 )} ( z , \overline{z} )$} (see the discussion after (\ref{eq:EIG06})). Away from $z = 0$, however, this term is irrelevant, and only the square root in (\ref{eq:EIG40}) contributes to the density --- \ieNotAPPB altogether,
\begin{equation}\label{eq:EIG41}
\rho_{\mathbf{P}} ( z , \overline{z} ) = \left\{ \begin{array}{ll} \frac{1}{\pi \sigma^{2}} \frac{R}{\sqrt{( 1 - R )^{2} + 4 R \frac{| z |^{2}}{\sigma^{2}}}} + f \delta^{( 2 )} ( z , \overline{z} ) , & \quad \textrm{for} \quad | z | \leq \sigma , \\ 0 , & \quad \textrm{for} \quad | z | > \sigma . \end{array} \right.
\end{equation}

As for any $L$, this function (\ref{eq:EIG41}) possesses rotational symmetry. One may check that it is correctly normalized, \smash{$\int_{0}^{\sigma} \dd r 2 \pi r \rho_{\mathbf{P}} ( z , \overline{z} ) |_{| z | = r} = 1$}, the delta function playing a vital role in this agreement. Setting $R = 1$ gives \smash{$\rho_{\mathbf{P}} ( z , \overline{z} ) = 1 / ( 2 \pi \sigma | z | )$}, for $| z | \leq \sigma$, in concord with the finding (\ref{eq:SquareGGProductMeanSpectralDensity}) of~\cite{BurdaJanikWaclaw2010}.


\subsubsection{The Singular Behavior of the Mean Spectral Density at Zero}
\label{sss:EIGTheSingularBehaviorOfTheMeanSpectralDensityAtZero}

Let us now examine one property of the mean spectral density of $\mathbf{P}$, namely, its behavior at the origin of the complex plane. We will show that the density is singular there, and that the rate of this explosion is governed by the number $s = 1 , 2 , \ldots , L$ (\ref{eq:s}) of the rectangularity ratios \smash{$R_{l}$} equal to $1$.

Indeed, we begin from recasting the main formula (\ref{eq:EIG37}) in the language of the non--holomorphic Green's function (\ref{eq:NonHolomorphicMTransformDefinition}), and explicitly distinguishing the terms with \smash{$R_{l} = 1$} from those with \smash{$R_{l} \neq 1$},
\begin{equation}\label{eq:EIG42}
z^{s} G_{\mathbf{P}} ( z , \overline{z} )^{s} \prod_{\substack{l \in \left\{ 1 , \ldots , L \right\} :\\R_{l} \neq 1}} \left( \frac{z G_{\mathbf{P}} ( z , \overline{z} )}{R_{l}} + 1 - \frac{1}{R_{l}} \right) = \frac{| z |^{2}}{\sigma^{2}} .
\end{equation}
Consider now how (\ref{eq:EIG42}) behaves in the limit of $z \to 0$. Let us suppose that as $z \to 0$, so also \smash{$z G_{\mathbf{P}} ( z , \overline{z} ) \to 0$}; we will justify this assumption \emph{a posteriori}. Consequently, the entire product over $l$ such that \smash{$R_{l} \neq 1$} in (\ref{eq:EIG42}) tends to a non--zero constant, and therefore, the singular behavior of the non--holomorphic Green's function reads
\begin{equation}\label{eq:EIG43}
z^{s} G_{\mathbf{P}} ( z , \overline{z} )^{s} \sim z \overline{z} , \quad \textrm{\ieNotAPPB} \quad G_{\mathbf{P}} ( z , \overline{z} ) \sim z^{\frac{1}{s} - 1} \overline{z}^{\frac{1}{s}} , \quad \textrm{as} \quad z \to 0 .
\end{equation}
(We confirm that our previous assumption holds, \smash{$z G_{\mathbf{P}} ( z , \overline{z} ) \sim | z |^{2 / s}$}.) Taking the derivative w.r.t. \smash{$\overline{z}$} of both sides of (\ref{eq:EIG43}), the singular behavior of the density (\ref{eq:MeanSpectralDensityFromNonHolomorphicGreensFunction}) is finally found,
\begin{equation}\label{eq:EIG44}
\rho_{\mathbf{P}} ( z , \overline{z} ) \sim | z |^{- 2 \left( 1 - \frac{1}{s} \right)} , \quad \textrm{as} \quad z \to 0 ,
\end{equation}
as anticipated in (\ref{eq:RhoPSingularityAtZero}). (Actually, for $s = 1$, and only for this value, there is no singularity, \smash{$\rho_{\mathbf{P}} ( z , \overline{z} ) \to \mathrm{const}$}.)


\subsubsection{The Behavior of the Mean Spectral Density at the Borderline}
\label{sss:EIGTheBehaviorOfTheMeanSpectralDensityAtTheBorderline}

Let us also focus for a while on the behavior of the density \smash{$\rho_{\mathbf{P}} ( z , \overline{z} )$} at the opposite end of the spectrum, namely, at the centered circle of radius $\sigma$, constituting the borderline of the average eigenvalues' domain. It has there a finite value of
\begin{equation}\label{eq:EIG45}
\rho_{\mathbf{P}} ( z , \overline{z} ) \left|_{| z | = \sigma} \right. = \frac{1}{\pi \sigma^{2}} R_{\mathrm{h}} , \quad \textrm{where} \quad \frac{1}{R_{\mathrm{h}}} \equiv \sum_{l = 1}^{L} \frac{1}{R_{l}} .
\end{equation}
(Note that it complies with (\ref{eq:EIG41}) and (\ref{eq:SquareGGProductMeanSpectralDensity}).) To see this, recall that for the rotationally--symmetric non--holomorphic $M$--transform, \smash{$M_{\mathbf{P}} ( z , \overline{z} ) = \mathfrak{M}_{\mathbf{P}} ( | z |^{2} )$}, the density is related to it as \smash{$\rho_{\mathbf{P}} ( z , \overline{z} ) = ( 1 / \pi ) \partial_{| z |^{2}} \mathfrak{M}_{\mathbf{P}} ( | z |^{2} )$}. Taking now the derivative w.r.t. \smash{$M_{\mathbf{P}} ( z , \overline{z} )$} of both sides of our main equation (\ref{eq:EIG37}), and setting \smash{$M_{\mathbf{P}} ( z , \overline{z} ) = 0$} (\ieNotAPPB the value at the boundary), we readily arrive at (\ref{eq:EIG45}).

Therefore in the considered thermodynamic limit (\ref{eq:ThermodynamicLimit}) the density has a jump from the value (\ref{eq:EIG45}) to zero as one crosses the frontier $| z | = \sigma$. However, for finite sizes of the random matrices in question, this step gets smoothed out. Imitating the idea of~\cite{ForresterHonner1999,Kanzieper2005,KhoruzhenkoSommers2009}, we propose the following model for this finite--$N$ (where by $N$ we denote the order of magnitude of the dimensions of the matrices, say \smash{$N \equiv N_{1}$}) smooth crossover from within the eigenvalues' circle to its outside: Since the mean spectral density of $\mathbf{P}$ is rotationally symmetric, it is sufficient to consider its radial part,
\begin{equation}\label{eq:EIG46}
\rho_{\mathbf{P}}^{\textrm{rad.}} ( r ) \equiv 2 \pi r \rho_{\mathbf{P}} ( z , \overline{z} ) \left|_{| z | = r} \right. .
\end{equation}
Subsequently, we introduce the following ``effective'' radial density, which is supposed to properly incorporate the finite--$N$ behavior at the borderline,
\begin{equation}\label{eq:EIG47}
\rho_{\mathbf{P}}^{\textrm{eff.}} ( r ) \equiv \rho_{\mathbf{P}}^{\textrm{rad.}} ( r ) \frac{1}{2} \mathrm{erfc} \left( q ( r - 1 ) \sqrt{N} \right) ,
\end{equation}
where the complementary error function is \smash{$\mathrm{erfc} ( x ) \equiv ( 2 / \sqrt{\pi} ) \int_{x}^{\infty} \dd t \exp ( - t^{2} )$}, while $q$ is a free parameter, whose value is to be adjusted by fitting. We numerically verify this hypothesis (\ref{eq:EIG47}) in the next paragraph.


\subsubsection{Numerical Confirmation}
\label{sss:EIGNumericalConfirmation}

We have tested the main equation (\ref{eq:EIG37}), as well as the finite--size ansatz (\ref{eq:EIG47}), quite extensively, obtaining excellent confirmation in all the considered situations, see figures~\ref{fig:EIGL2a}, \ref{fig:EIGL2b}, \ref{fig:EIGL2c}, \ref{fig:EIGL2d} and~\ref{fig:EIGL34}.


\section{The Singular Values of a Product of Rectangular Gaussian Random Matrices}
\label{s:TheSingularValuesOfAProductOfRectangularGaussianRandomMatrices}


\subsection{Derivation}
\label{ss:SVDerivation}

This subsection is devoted to deriving the second main formula of this article, (\ref{eq:MQBasicEquation}), \ieNotAPPB an $( L + 1 )$--th--order polynomial equation obeyed by the $M$--transform, a quantity equivalent to the average density of the eigenvalues, of the Hermitian matrix \smash{$\mathbf{Q} \equiv \mathbf{P}^{\dagger} \mathbf{P}$} (\ref{eq:SingularValuesDefinition}), with $\mathbf{P}$ the product (\ref{eq:ProductDefinition}) of rectangular (\ref{eq:ThermodynamicLimit}) Gaussian random matrices (\ref{eq:RectangularGGMeasure}). The underlying idea will be to rewrite $\mathbf{Q}$ as a product of some Hermitian matrices, to which the FRV (appendix~\ref{s:FreeRandomVariablesInANutshell}) multiplication law (\ref{eq:NonCommutativeMultiplicationLaw}) will be harnessed.

Let us commence from defining, for any $l = 1 , 2 , \ldots , L$, a square \smash{$N_{l + 1} \times N_{l + 1}$} matrix
$$
\mathbf{Q}_{l} \equiv \left( \mathbf{A}_{1} \mathbf{A}_{2} \ldots \mathbf{A}_{l - 1} \mathbf{A}_{l} \right)^{\dagger} \left( \mathbf{A}_{1} \mathbf{A}_{2} \ldots \mathbf{A}_{l - 1} \mathbf{A}_{l} \right) =
$$
\begin{equation}\label{eq:SV01}
= \mathbf{A}_{l}^{\dagger} \mathbf{A}_{l - 1}^{\dagger} \ldots \mathbf{A}_{2}^{\dagger} \mathbf{A}_{1}^{\dagger} \mathbf{A}_{1} \mathbf{A}_{2} \ldots \mathbf{A}_{l - 1} \mathbf{A}_{l} ,
\end{equation}
being a generalization of $\mathbf{Q}$ which includes only the first $l$ random matrices, as well as a square \smash{$N_{l} \times N_{l}$} matrix, which differs from \smash{$\mathbf{Q}_{l}$} only by the position of the last matrix in the string, \ieNotAPPB \smash{$\mathbf{A}_{l}$}, which is now placed as the first matrix in the string,
\begin{equation}\label{eq:SV02}
\widetilde{\mathbf{Q}}_{l} \equiv \mathbf{A}_{l} \mathbf{A}_{l}^{\dagger} \mathbf{A}_{l - 1}^{\dagger} \ldots \mathbf{A}_{2}^{\dagger} \mathbf{A}_{1}^{\dagger} \mathbf{A}_{1} \mathbf{A}_{2} \ldots \mathbf{A}_{l - 1} = \left( \mathbf{A}_{l} \mathbf{A}_{l}^{\dagger} \right) \mathbf{Q}_{l - 1} .
\end{equation}
We are interested in the eigenvalues of the Hermitian matrix \smash{$\mathbf{Q} = \mathbf{Q}_{L}$}.

The orders of the terms in the two above products (\ref{eq:SV01}), (\ref{eq:SV02}) are related to each other by a cyclic shift, therefore, for any integer $n \geq 1$, there will be \smash{$\Tr \mathbf{Q}_{l}^{n} = \Tr \widetilde{\mathbf{Q}}_{l}^{n}$}. Hence, the $M$--transforms (\ref{eq:MTransformExpansion}) of the two above random matrices are related by
\begin{equation}\label{eq:SV03}
M_{\mathbf{Q}_{l}} ( z ) = \sum_{n \geq 1} \frac{1}{z^{n}} \frac{1}{N_{l + 1}} \la \Tr \mathbf{Q}_{l}^{n} \ra = \frac{N_{l}}{N_{l + 1}} \sum_{n \geq 1} \frac{1}{z^{n}} \frac{1}{N_{l}} \la \Tr \widetilde{\mathbf{Q}}_{l}^{n} \ra = \frac{R_{l}}{R_{l + 1}} M_{\widetilde{\mathbf{Q}}_{l}} ( z ) .
\end{equation}
Inverting (\ref{eq:SV03}) functionally, we obtain a relationship between the respective $N$--transforms (\ref{eq:NTransformDefinition}),
\begin{equation}\label{eq:SV04}
N_{\mathbf{Q}_{l}} ( z ) = N_{\widetilde{\mathbf{Q}}_{l}} \left( \frac{R_{l + 1}}{R_{l}} z \right) .
\end{equation}

The reason for introducing the auxiliary matrix \smash{$\widetilde{\mathbf{Q}}_{l}$} (\ref{eq:SV02}) is that it is a product of two free matrices, \smash{$\mathbf{A}_{l} \mathbf{A}_{l}^{\dagger}$} and \smash{$\mathbf{Q}_{l - 1}$}. Using the FRV multiplication law (\ref{eq:NonCommutativeMultiplicationLaw}), we can thus write, for all $l = 2 , 3 , \ldots , L$,
\begin{equation}\label{eq:SV05}
N_{\widetilde{\mathbf{Q}}_{l}} ( z ) = \frac{z}{z + 1} N_{\mathbf{A}_{l} \mathbf{A}_{l}^{\dagger}} ( z ) N_{\mathbf{Q}_{l - 1}} ( z ) .
\end{equation}

{}From these two equations (\ref{eq:SV04}), (\ref{eq:SV05}), we now eliminate the $N$--transform of the auxiliary \smash{$\widetilde{\mathbf{Q}}_{l}$}, leaving us with the following recurrence relation for the $N$--transform of \smash{$\mathbf{Q}_{l}$},
\begin{equation}\label{eq:SV06}
N_{\mathbf{Q}_{l}} ( z ) = \frac{z}{z + \frac{R_{l}}{R_{l + 1}}} N_{\mathbf{A}_{l} \mathbf{A}_{l}^{\dagger}} \left( \frac{R_{l + 1}}{R_{l}} z \right) N_{\mathbf{Q}_{l - 1}} \left( \frac{R_{l + 1}}{R_{l}} z \right) , \quad \textrm{for} \quad l = 2 , 3 , \ldots , L ,
\end{equation}
with the initial condition,
\begin{equation}\label{eq:SV07}
N_{\mathbf{Q}_{1}} ( z ) = N_{\widetilde{\mathbf{Q}}_{1}} \left( \frac{R_{2}}{R_{1}} z \right) = N_{\mathbf{A}_{1} \mathbf{A}_{1}^{\dagger}} \left( \frac{R_{2}}{R_{1}} z \right) ,
\end{equation}
which stems from (\ref{eq:SV04}) and the fact that \smash{$\widetilde{\mathbf{Q}}_{1} = \mathbf{A}_{1} \mathbf{A}_{1}^{\dagger}$}. The solution of this recurrence (\ref{eq:SV06}), (\ref{eq:SV07}) is then readily found to be
$$
N_{\mathbf{Q}_{L}} ( z ) = \frac{z^{L - 1}}{\left( z + R_{2} \right) \left( z + R_{3} \right) \ldots \left( z + R_{L} \right)} \cdot
$$
\begin{equation}\label{eq:SV08}
\cdot N_{\mathbf{A}_{1} \mathbf{A}_{1}^{\dagger}} \left( \frac{z}{R_{1}} \right) N_{\mathbf{A}_{2} \mathbf{A}_{2}^{\dagger}} \left( \frac{z}{R_{2}} \right) \ldots N_{\mathbf{A}_{L} \mathbf{A}_{L}^{\dagger}} \left( \frac{z}{R_{L}} \right) .
\end{equation}

It remains now to find the $N$--transforms of the random matrices \smash{$\mathbf{A}_{l} \mathbf{A}_{l}^{\dagger}$}. They are examples of the so--called ``Wishart ensembles,'' and the problem of computing their $N$--transforms, with the precise normalization of the probability measures of the \smash{$\mathbf{A}_{l}$}'s which we are employing (\ref{eq:RectangularGGMeasure}), has first been solved in~\cite{FeinbergZee1997}: Expressions (1.8), (2.8), (2.13), (2.14) of this article yield the Green's function of \smash{$\mathbf{A}_{l} \mathbf{A}_{l}^{\dagger}$}, which immediately leads to the pertinent $N$--transform,
\begin{equation}\label{eq:SV09}
N_{\mathbf{A}_{l} \mathbf{A}_{l}^{\dagger}} ( z ) = \sigma_{l}^{2} \frac{( z + 1 ) \left( \sqrt{\frac{N_{l}}{N_{l + 1}}} z + \sqrt{\frac{N_{l + 1}}{N_{l}}} \right)}{z} .
\end{equation}
Substituting (\ref{eq:SV09}) into (\ref{eq:SV08}), one finally arrives at the desired formula for the $N$--transform of \smash{$\mathbf{Q} = \mathbf{Q}_{L}$},
\begin{equation}\label{eq:SV10}
N_{\mathbf{Q}} ( z ) = \sigma^{2} \sqrt{R_{1}} \frac{1}{z} ( z + 1 ) \left( \frac{z}{R_{1}} + 1 \right) \left( \frac{z}{R_{2}} + 1 \right) \ldots \left( \frac{z}{R_{L}} + 1 \right) ,
\end{equation}
with $\sigma$ defined in (\ref{eq:Sigma}). In other words, the corresponding $M$--transform \smash{$M_{\mathbf{Q}} ( z )$} satisfies the following polynomial equation of order $( L + 1 )$,
\begin{equation}\label{eq:SV11}
\sqrt{R_{1}} \frac{1}{M_{\mathbf{Q}} ( z )} \left( M_{\mathbf{Q}} ( z ) + 1 \right) \left( \frac{M_{\mathbf{Q}} ( z )}{R_{1}} + 1 \right) \left( \frac{M_{\mathbf{Q}} ( z )}{R_{2}} + 1 \right) \ldots \left( \frac{M_{\mathbf{Q}} ( z )}{R_{L}} + 1 \right) = \frac{z}{\sigma^{2}} ,
\end{equation}
or in the case of \smash{$N_{L + 1} = N_{1}$} (\ieNotAPPB \smash{$R_{1} = 1$}, required when one wishes for $\mathbf{P}$ to have eigenvalues too),
\begin{equation}\label{eq:SV12}
\frac{1}{M_{\mathbf{Q}} ( z )} \left( M_{\mathbf{Q}} ( z ) + 1 \right)^{2} \left( \frac{M_{\mathbf{Q}} ( z )}{R_{2}} + 1 \right) \ldots \left( \frac{M_{\mathbf{Q}} ( z )}{R_{L}} + 1 \right) = \frac{z}{\sigma^{2}} .
\end{equation}
This completes our derivation of (\ref{eq:MQBasicEquation}).

\begin{figure}[ht]
\begin{center}
\includegraphics[width=8.5cm]{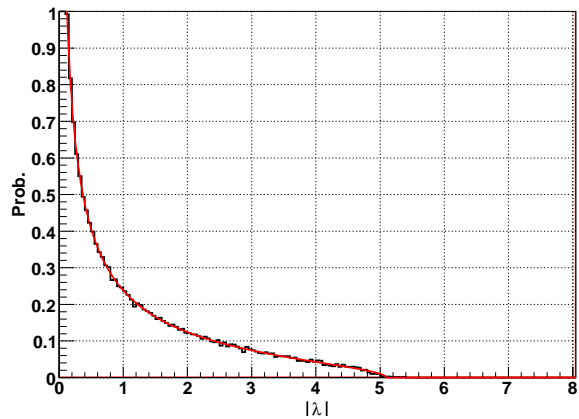}
\caption{Numerical verification of the theoretical formula (\ref{eq:SV11}) for the mean spectral density \smash{$\rho_{\mathbf{Q}} ( \lambda )$} of the random matrix \smash{$\mathbf{Q} = \mathbf{P}^{\dagger} \mathbf{P}$} (\ref{eq:SingularValuesDefinition}). Everywhere \smash{$N_{L + 1} = N_{1} = 50$}. The number of Monte--Carlo iterations is $20,\!000$, \ieNotAPPB all the histograms are generated from \smash{$10^{6}$} eigenvalues. Here $L = 2$, and the matrix sizes are chosen to be \smash{$N_{1} = 50$}, \smash{$N_{2} = 150$}.}
\label{fig:SVL234a}
\end{center}
\end{figure}

\begin{figure}[ht]
\begin{center}
\includegraphics[width=8.5cm]{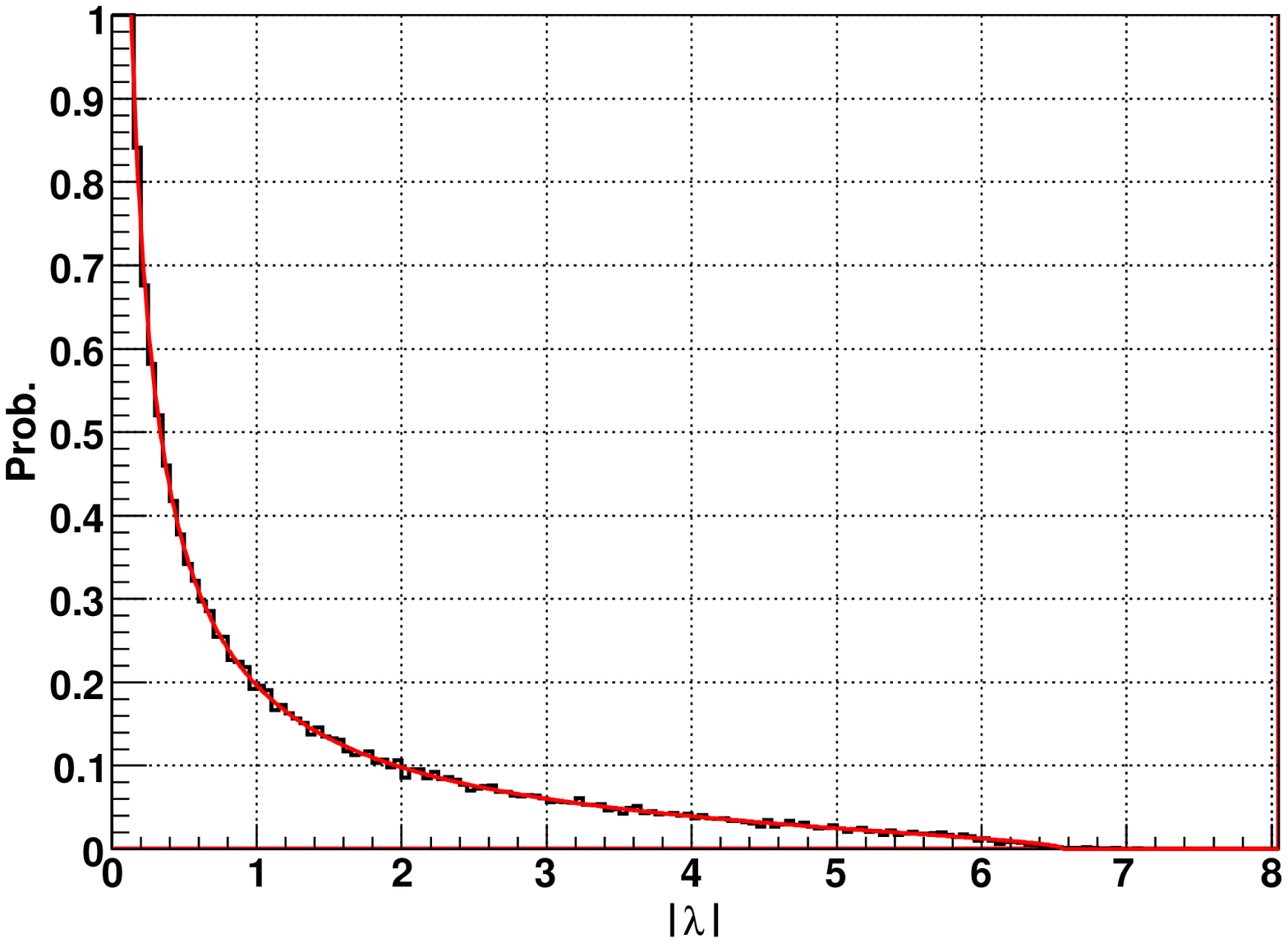}
\caption{The same as figure~\ref{fig:SVL234a}, but with $L = 3$, and the matrix sizes are \smash{$N_{1} = 50$}, \smash{$N_{2} = 100$}, \smash{$N_{3} = 150$}.}
\label{fig:SVL234b}
\end{center}
\end{figure}

\begin{figure}[ht]
\begin{center}
\includegraphics[width=8.5cm]{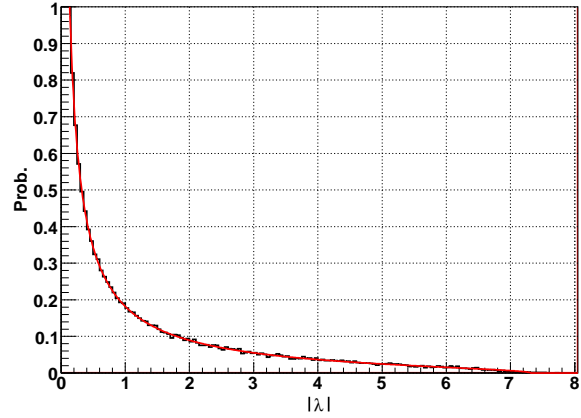}
\caption{The same as figure~\ref{fig:SVL234a}, but with $L = 4$, and the matrix sizes are \smash{$N_{1} = 50$}, \smash{$N_{2} = 100$}, \smash{$N_{3} = 150$}, \smash{$N_{4} = 200$}.}
\label{fig:SVL234c}
\end{center}
\end{figure}


\subsection{Comments}
\label{ss:SVComments}


\subsubsection{The Singular Behavior of the Mean Spectral Density at Zero}
\label{sss:SVTheSingularBehaviorOfTheMeanSpectralDensityAtZero}

Following an analogous line of reasoning as in \S\ref{sss:EIGTheSingularBehaviorOfTheMeanSpectralDensityAtZero}, we may unravel the singular behavior at zero of the mean spectral density of $\mathbf{Q}$ stemming from the main formula (\ref{eq:SV11}). Again, we rewrite this equation in terms of the (usual, holomorphic) Green's function (\ref{eq:MTransformDefinition}), and separate the terms with \smash{$R_{l} = 1$} (the number of such rectangularity ratios is called $s$ (\ref{eq:s})) from those with \smash{$R_{l} \neq 1$},
\begin{equation}\label{eq:SV13}
\frac{1}{z G_{\mathbf{Q}} ( z ) - 1} z^{s + 1} G_{\mathbf{Q}} ( z )^{s + 1} \prod_{\substack{l \in \left\{ 1 , \ldots , L \right\} :\\R_{l} \neq 1}} \left( \frac{z G_{\mathbf{Q}} ( z )}{R_{l}} + 1 - \frac{1}{R_{l}} \right) = \frac{z}{\sigma^{2}} .
\end{equation}
In the limit $z \to 0$, the relevant terms in (\ref{eq:SV13}) will thus read
\begin{equation}\label{eq:SV14}
z^{s + 1} G_{\mathbf{Q}} ( z )^{s + 1} \sim z , \quad \textrm{\ieNotAPPB} \quad G_{\mathbf{Q}} ( z ) \sim z^{- \frac{s}{s + 1}} , \quad \textrm{as} \quad z \to 0 ,
\end{equation}
which immediately leads to the aforementioned result (\ref{eq:RhoQSingularityAtZero}).


\subsubsection{Numerical Confirmation}
\label{sss:SVNumericalConfirmation}

We have performed several numerical tests of the main formula (\ref{eq:SV11}), in all cases obtaining astonishing agreement, see figures~\ref{fig:SVL234a}, \ref{fig:SVL234b} and~\ref{fig:SVL234c}.


\section{Conclusions}
\label{s:Conclusions}


\subsection{Summary}
\label{ss:Summary}

The main contribution of this article is given by the two equations (\ref{eq:MPBasicEquation}) and (\ref{eq:MQBasicEquation}) for the $M$--transforms, \ieNotAPPB objects conveniently encoding the average spectral densities, of the product \smash{$\mathbf{P} = \mathbf{A}_{1} \mathbf{A}_{2} \ldots \mathbf{A}_{L}$} (\ref{eq:ProductDefinition}) of an arbitrary number $L$ of independent rectangular (\ref{eq:ThermodynamicLimit}) Gaussian random matrices (\ref{eq:RectangularGGMeasure}), as well as of the matrix \smash{$\mathbf{Q} = \mathbf{P}^{\dagger} \mathbf{P}$} (\ref{eq:SingularValuesDefinition}), whose eigenvalues are the squared singular values of $\mathbf{P}$, respectively. Both these equations are polynomial, and have orders $L$ and $( L + 1 )$, respectively, so in general they may be solved only numerically; however, some properties of the mean spectral densities can still be retrieved analytically, such as their singular behavior at zero (\ref{eq:RhoPSingularityAtZero}), (\ref{eq:RhoQSingularityAtZero}).

Furthermore, (\ref{eq:MPBasicEquation}) and (\ref{eq:MQBasicEquation}) are very similar to each other --- the fact which directed us to put forward a conjecture (\ref{eq:Conjecture}) that the same resemblance is a feature shared by all non--Hermitian ensembles $\mathbf{X}$ possessing rotationally--symmetric average distribution of the eigenvalues. For such models, the non--holomorphic $M$--transform \smash{$M_{\mathbf{X}} ( z , \overline{z} )$} (\ref{eq:NonHolomorphicMTransformDefinition}) is a function of the real argument \smash{$| z |^{2}$} (\ref{eq:MPAsAFunctionOfTheRadius}), thereby allowing functional inversion, and hence, a definition of the ``rotationally--symmetric non--holomorphic $N$--transform'' (\ref{eq:RotationallySymmetricNonHolomorphicNTransformDefinition}) --- even though for general non--Hermitian random matrices a construction of a ``non--holomorphic $N$--transform'' remains thus far unknown. This new $N$--transform is then conjectured to be in a simple relationship (\ref{eq:Conjecture}) to the (usual) $N$--transform of the Hermitian ensemble \smash{$\mathbf{X}^{\dagger} \mathbf{X}$}. In a typical situation, the latter will be solvable much more easily than the former, owing to the plethora of tools devised in the Hermitian world, albeit the opposite may be true as well. This is indeed the case here --- our derivation of (\ref{eq:MPBasicEquation}), based on non--Hermitian planar diagrammatics and Dyson--Schwinger's equations, is much more involved than a simple application of the FRV multiplication rule leading to (\ref{eq:MQBasicEquation}) --- and consequently, the hypothesis (\ref{eq:Conjecture}) would provide a means of avoiding the complicated diagrammatics. To the best of our knowledge, this would be the first use of the free random variables calculus to computing the mean spectral density of a non--Hermitian product of random matrices.

We have also suggested a model for a finite--size behavior of the density of $\mathbf{P}$ near the borderline of the eigenvalues' domain (\ref{eq:EIG47}), taking after the work of~\cite{ForresterHonner1999,Kanzieper2005,KhoruzhenkoSommers2009}. It performs outstandingly well when checked against numerical simulations.


\subsection{Possible Applications}
\label{ss:PossibleApplications}

Let us now succinctly sketch some possible applications of these results to wireless telecommunication, quantum entanglement and finance.


\subsubsection{Wireless Telecommunication}
\label{sss:WirelessTelecommunication}

Information theory for wireless telecommunication has been intensively developed in the past decade, after it had been realized that the information rate can be increased by an introduction of multiple antenna channels, known as the ``multiple--input, multiple--output'' (MIMO) transmission links. The MIMO capacity for Gaussian channels has been calculated in the pioneering work~\cite{Telatar1999}, triggering large activity in the field. Immediately, it became clear that an appropriate language and methods to address this type of problems are provided by random matrix theory (consult~\cite{TulinoVerdu2004} for a review).

The MIMO capacity reads~\cite{Telatar1999},
\begin{equation}\label{eq:MIMOCapacity}
\textrm{Capacity } C = \la \log_{2} \Det \left( \Id_{N_{\textrm{rec.}}} + \frac{\mathrm{SNR}}{N_{\textrm{tr.}}} \mathbf{A} \mathbf{C} \mathbf{A}^{\dagger} \right) \ra ,
\end{equation}
where \smash{$N_{\textrm{rec.}}$} is the number of receivers, \smash{$N_{\textrm{tr.}}$} of transmitters, and $\mathrm{SNR}$ the signal--to--noise ratio. The output signals \smash{$\vec{y}$} are calculated from the input \smash{$\vec{x}$} as
\begin{equation}\label{eq:MIMOSystem}
\vec{y} = \sqrt{\frac{\mathrm{SNR}}{N_{\textrm{tr.}}}} \mathbf{A} \vec{x} + \vec{\eta} ,
\end{equation}
where $\mathbf{A}$ is the response matrix for a given frequency, \smash{$\vec{\eta}$} is a standardized multivariate white noise, while $\mathbf{C}$ is a covariance matrix for the input signals, \ieNotAPPB \smash{$[ \mathbf{C} ]_{a b} \equiv \langle x_{a} x_{b} \rangle$}.

In the simplest case, one assumes that the input signals are uncorrelated, \ieNotAPPB \smash{$\mathbf{C} = \Id_{N_{\textrm{tr.}}}$}, and that $\mathbf{A}$ is a random matrix built of IID centered Gaussian entries. This corresponds to a random situation, when one has no information about the signal propagation. Then, the asymptotic (\smash{$N_{\textrm{tr.}} \to \infty$}) mutual information per channel can be derived to be \smash{$\mu = \int \dd \lambda \rho_{\mathbf{Q}} ( \lambda ) \log ( 1 + \mathrm{SNR} \lambda )$}, where \smash{$\rho_{\mathbf{Q}} ( \lambda )$} is the limiting mean eigenvalue density of the Wishart matrix \smash{$\mathbf{Q} = \mathbf{A}^{\dagger} \mathbf{A}$}~\cite{TulinoVerdu2004}.

The model considered in our paper can be applied to a more complex case of signals traveling over $L$ consecutive MIMO links~\cite{Muller2002}: It is first sent from \smash{$N_{1}$} transmitters via a MIMO link to \smash{$N_{2}$} receivers, which then re--transmit it via a new MIMO link to the subsequent \smash{$N_{3}$} receivers, \emph{etc.} Clearly, the capacity will depend on these numbers of intermediate re--transmitters; in particular, if any of the \smash{$N_{l}$}'s is small, the capacity will be reduced. The output now reads
\begin{equation}\label{eq:MIMOSystemWithLReceivers}
\vec{y} = \sqrt{\frac{\mathrm{SNR}}{N_{1}}} \mathbf{A}_{L} \ldots \mathbf{A}_{2} \mathbf{A}_{1} \vec{x} + \vec{\eta} ,
\end{equation}
and therefore, the effective propagation is given by the matrix \smash{$\mathbf{P} = \mathbf{A}_{L} \ldots \mathbf{A}_{2} \mathbf{A}_{1}$}, whereas the mutual information per channel by \smash{$\mu_{L} = \int \dd \lambda \rho_{\mathbf{Q}} ( \lambda ) \log ( 1 + \mathrm{SNR} \lambda )$}, with \smash{$\mathbf{Q} = \mathbf{P}^{\dagger} \mathbf{P}$}. Now, it is precisely our second main result (\ref{eq:MQBasicEquation}) which can be exploited to find the relevant density \smash{$\rho_{\mathbf{Q}} ( \lambda )$}.

Let us finally just mention that one could imagine a more general situation, where MIMO links form a directed network --- each directed link $l m$ representing a single MIMO channel between \smash{$N_{l}$} transmitters and \smash{$N_{m}$} receivers. (The discussed case (\ref{eq:MIMOSystemWithLReceivers}) corresponds to a linear graph, $1 \to 2 \to \ldots \to L$.)


\subsubsection{Quantum Entanglement}
\label{sss:QuantumEntanglement}

A complex directed network of MIMO links is somewhat similar to the structures appearing in the context of quantum entanglement. There, one considers graphs whose edges describe bi--partite maximally entangled states, while vertices --- the couplings between subsystems residing at the same vertex~\cite{CollinsNechitaZyczkowski2010}. In the simplest case of a graph consisting of a single link, it is just a bi--partite entangled state. The corresponding density matrix for a bi--partite subsystem is given by \smash{$\mathbf{Q} = \mathbf{A}^{\dagger} \mathbf{A}$}, where $\mathbf{A}$ is a rectangular \smash{$N_{1} \times N_{2}$} matrix defining a pure state, \smash{$| \psi \rangle = \sum_{a = 1}^{N_{1}} \sum_{b = 1}^{N_{2}} [ \mathbf{A} ]_{a b} | \alpha_{a} \rangle \otimes | \beta_{b} \rangle$}, being a combination of the basis states in the subsystem, \smash{$| \alpha_{a} \rangle$} and \smash{$| \beta_{b} \rangle$} (see for instance~\cite{Majumdar2010}). One can easily find that linear graphs with additional loops at the end vertices correspond. The density matrix for the subsystem sitting in the end vertex is given by \smash{$\mathbf{Q} = \mathbf{P}^{\dagger} \mathbf{P}$}, where \smash{$\mathbf{P} = \mathbf{A}_{1} \mathbf{A}_{2} \ldots \mathbf{A}_{L}$}~\cite{CollinsNechitaZyczkowski2010}. If all the subsystems are of the same size, the average spectral distributions of $\mathbf{Q}$ are known~\cite{BanicaBelinschiCapitaineCollins2007,BenaychGeorges2008} as the ``Fuss--Catalan family''~\cite{Armstrong2009}; they can be obtained from (\ref{eq:MQBasicEquation}) by setting all the \smash{$R_{l}$}'s to $1$. However, if the subsystems have different sizes, one needs to apply our general formula (\ref{eq:MQBasicEquation}).


\subsubsection{Finance}
\label{sss:Finances}

Another area of applications is financial engineering. Products of rectangular random matrices may be used in calculations of lagged correlation functions, which play very important roles in risk management and portfolio theory. This issue will be discussed in a forthcoming publication.


\section*{Acknowledgements}

We would like to thank Romuald A. Janik, Bart{\l}omiej Wac{\l}aw and Karol \.{Z}yczkowski for interesting discussions.

This work was partially supported by the Polish Ministry of Science and Higher Education Grants: No.~N~N202~229137~(2009--2012) and ``Iuventus Plus'' No.~0148/H03/2010/70. AJ acknowledges the support of Clico Ltd., Oleandry 2, 30--063 Krak\'{o}w, Poland, while completing parts of this paper.


\appendix

\section{The Diagrammatic Approach to Solving Matrix Models}
\label{s:TheDiagrammaticApproachToSolvingMatrixModels}


\subsection{\ Hermitian Random Matrices}
\label{ss:HermitianRandomMatrices}


\subsubsection{\ The Green's Function and $M$--Transform}
\label{sss:TheGreensFunctionAndMTransform}

The most basic question one asks in RMT about an $N \times N$ --- where we take the thermodynamic limit of $N \to \infty$ --- Hermitian random matrix $\mathbf{H}$, endowed with some probability measure $\dd \mu ( \mathbf{H} )$, is to compute the average distribution of its (real) eigenvalues \smash{$\lambda_{1} , \lambda_{2} , \ldots , \lambda_{N}$}, defined as
\begin{equation}\label{eq:MeanSpectralDensityDefinition}
\rho_{\mathbf{H}} ( \lambda ) \equiv \frac{1}{N} \sum_{a = 1}^{N} \la \delta \left( \lambda - \lambda_{a} \right) \ra ,
\end{equation}
where the average map $\langle \ldots \rangle \equiv \int ( \ldots ) \dd \mu ( \mathbf{H} )$, while $\delta ( \lambda )$ denotes the real Dirac's delta function.

However, it proves more convenient to work with another equivalent object --- the ``Green's function'' (a.k.a. ``resolvent''). First, one introduces the $N \times N$ ``Green's function matrix,''
\begin{equation}\label{eq:GreensFunctionMatrixDefinition}
\mathbf{G}_{\mathbf{H}} ( z ) \equiv \la ( \mathbf{Z} - \mathbf{H} )^{- 1} \ra , \quad \textrm{where} \quad \mathbf{Z} \equiv z \Id_{N} ,
\end{equation}
where $z$ is a complex argument, and \smash{$\Id_{N}$} represents the $N \times N$ unit matrix. Then, the Green's function is defined as its normalized trace,
\begin{equation}\label{eq:GreensFunctionDefinition}
G_{\mathbf{H}} ( z ) \equiv \frac{1}{N} \Tr \mathbf{G}_{\mathbf{H}} ( z ) = \frac{1}{N} \sum_{a = 1}^{N} \la \frac{1}{z - \lambda_{a}} \ra .
\end{equation}
As clearly seen in this definition, \smash{$G_{\mathbf{H}} ( z )$} is, for any finite $N$, a meromorphic function, with poles located at the average spectrum. When $N$ tends to infinity, these poles coalesce into continuous intervals on the real axis, and \smash{$G_{\mathbf{H}} ( z )$} turns into a function holomorphic everywhere on the complex plane excluding the cuts formed by these intervals. Knowing the mean density of the eigenvalues in these cuts allows one to reproduce the Green's function,
\begin{equation}\label{eq:GreensFunctionFromMeanSpectralDensity}
G_{\mathbf{H}} ( z ) = \int_{\mathrm{cuts}} \dd \lambda \rho_{\mathbf{H}} ( \lambda ) \frac{1}{z - \lambda}
\end{equation}
(in other words, the Green's function is the Stjeltes' transform of the density), and conversely, a standard representation of the real Dirac's delta,
\begin{equation}\label{eq:RealDiracsDelta}
\delta ( \lambda ) = - \frac{1}{\pi} \lim_{\epsilon \to 0^{+}} \im \frac{1}{\lambda + \ii \epsilon}
\end{equation}
implies that the mean spectral density is recovered when one approaches with the argument of the Green's function the eigenvalues' cuts along the imaginary direction,
\begin{equation}\label{eq:MeanSpectralDensityFromGreensFunction}
\rho_{\mathbf{H}} ( \lambda ) = - \frac{1}{\pi} \lim_{\epsilon \to 0^{+}} \im G_{\mathbf{H}} ( \lambda + \ii \epsilon ) .
\end{equation}
We thus see that (\ref{eq:MeanSpectralDensityDefinition}) and (\ref{eq:GreensFunctionDefinition}) carry identical information.

Instead of the Green's function, one often prefers yet another quantity, the ``$M$--transform,'' simply related to the former,
\begin{equation}\label{eq:MTransformDefinition}
M_{\mathbf{H}} ( z ) \equiv z G_{\mathbf{H}} ( z ) - 1 .
\end{equation}
Its meaning is deduced from considering its large--$z$ expansion --- being holomorphic everywhere except the cuts on the real line, it permits a power--series expansion around $z = \infty$,
\begin{equation}\label{eq:MTransformExpansion}
M_{\mathbf{H}} ( z ) = \sum_{n \geq 1} \frac{M_{\mathbf{H} , n}}{z^{n}} , \quad \textrm{where} \quad M_{\mathbf{H} , n} \equiv \frac{1}{N} \Tr \la \mathbf{H}^{n} \ra = \int_{\mathrm{cuts}} \dd \lambda \rho_{\mathbf{H}} ( \lambda ) \lambda^{n} .
\end{equation}
The coefficients are precisely the moments of the probability distribution in question, and so, \smash{$M_{\mathbf{H}} ( z )$} is also known as the ``moments' generating function.''


\subsubsection{\ The Green's Function for the GUE from Planar Diagrams}
\label{sss:TheGreensFunctionForTheGUEFromPlanarDiagrams}

There are multiple techniques developed for the purpose of calculating the Green's function of a Hermitian random matrix $\mathbf{H}$, and we will now briefly describe the method of representing it as a sum of planar fat diagrams, which is then performed using the so--called ``Dyson--Schwinger's equations.'' We will not aim at a comprehensive presentation, but only show how the approach works on a simplest example of a Hermitian model --- the ``Gaussian Unitary Ensemble'' (GUE), with standard deviation $\sigma$,
\begin{equation}\label{eq:GUEMeasure}
\dd \mu \left( \mathbf{H} \right) \propto \ee^{- \frac{N}{2 \sigma^{2}} \Tr \mathbf{H}^{2}} \DD \mathbf{H} .
\end{equation}
(It means that the real elements on the diagonal are IID with the same variance \smash{$\sigma^{2} / N$}, while the real and imaginary parts of the entries below the diagonal are IID with the variance \smash{$\sigma^{2} / 2 N$}.)

One begins from expanding the Green's function matrix (\ref{eq:GreensFunctionMatrixDefinition}) into the power--series around infinite $z$,
\begin{equation}\label{eq:GreensFunctionMatrixExpansion}
\mathbf{G}_{\mathbf{H}} ( z ) = \mathbf{Z}^{- 1} + \la \mathbf{Z}^{- 1} \mathbf{H} \mathbf{Z}^{- 1} \mathbf{H} \mathbf{Z}^{- 1} \ra + \la \mathbf{Z}^{- 1} \mathbf{H} \mathbf{Z}^{- 1} \mathbf{H} \mathbf{Z}^{- 1} \mathbf{H} \mathbf{Z}^{- 1} \mathbf{H} \mathbf{Z}^{- 1} \ra + \ldots ,
\end{equation}
where we have preserved only even moments, as the probability measure (\ref{eq:GUEMeasure}) is even, as well as kept the order of the \smash{$\mathbf{Z}^{- 1}$}'s and $\mathbf{H}$'s intact, even though they commute. Since $\mathbf{Z}$ is independent of $\mathbf{H}$, and can thus be taken outside of the average, one is left at every order with an expectation of the form \smash{$\langle [ \mathbf{H} ]_{a_{1} a_{2}} [ \mathbf{H} ]_{a_{3} a_{4}} \ldots [ \mathbf{H} ]_{a_{2 n - 1} a_{2 n}} \rangle$}. Owing to Wick's theorem, any such $n$--point correlation function can be expressed --- by making all the possible contractions --- through the $2$--point correlation function (propagator), which for the GUE measure (\ref{eq:GUEMeasure}) reads
\begin{equation}\label{eq:GUEPropagator}
\la [ \mathbf{H} ]_{a b} [ \mathbf{H} ]_{c d} \ra = \frac{\sigma^{2}}{N} \delta_{a d} \delta_{b c} .
\end{equation}

Consequently, the structure of this series (\ref{eq:GreensFunctionMatrixExpansion}) lends itself to a graphical representation, in which every factor of \smash{$[ \mathbf{Z}^{- 1} ]_{a b}$} is depicted as a straight line connecting the matrix indices $a$ and $b$, while every propagator \smash{$\langle [ \mathbf{H} ]_{a b} [ \mathbf{H} ]_{c d} \rangle$} as a double arc connecting two pairs of indices, $a b$ and $c d$,

\setlength{\unitlength}{0.8pt}
\begin{picture}(500,82)(0,3)
\thicklines

\put(10,50){\circle*{6}}
\put(8,38){$a$}
\put(50,50){\circle*{6}}
\put(48,38){$b$}
\put(10,50){\line(1,0){40}}
\put(60,50){$= \left[ \mathbf{Z}^{- 1} \right]_{a b} = \frac{1}{z} \delta_{a b}$,}

\put(210,30){\circle*{6}}
\put(208,18){$a$}
\put(220,30){\circle*{6}}
\put(218,18){$b$}
\put(280,30){\circle*{6}}
\put(278,18){$c$}
\put(290,30){\circle*{6}}
\put(288,18){$d$}
\put(250,30){\arc{80}{3.1416}{6.2832}}
\put(250,30){\arc{60}{3.1416}{6.2832}}
\put(300,50){$= \la [ \mathbf{H} ]_{a b} [ \mathbf{H} ]_{c d} \ra = \frac{\sigma^{2}}{N} \delta_{a d} \delta_{b c}$.}
\end{picture}
The first three orders of (\ref{eq:GreensFunctionMatrixExpansion}) are thus drawn as

\begin{picture}(500,80)(0,415)
\thicklines

\put(10,450){\circle*{6}}
\put(8,438){$a$}
\put(90,450){\circle*{6}}
\put(88,438){$b$}
\put(10,450){\line(1,0){20}}
\put(90,450){\line(-1,0){20}}
\filltype{shade}
\put(50,450){\circle*{40}}
\filltype{black}
\put(46,447){{\color{white}$\mathbf{G}$}}
\put(30,415){$[ \mathbf{G}_{\mathbf{H}} ( z ) ]_{a b}$}

\put(102,450){$=$}

\put(120,450){\circle*{6}}
\put(118,438){$a$}
\put(160,450){\circle*{6}}
\put(158,438){$b$}
\put(120,450){\line(1,0){40}}

\put(172,450){$+$}

\put(190,450){\circle*{6}}
\put(188,438){$a$}
\put(230,450){\circle*{6}}
\put(227,438){$c_{1}$}
\put(240,450){\circle*{6}}
\put(237,438){$c_{2}$}
\put(280,450){\circle*{6}}
\put(277,438){$c_{3}$}
\put(290,450){\circle*{6}}
\put(287,438){$c_{4}$}
\put(330,450){\circle*{6}}
\put(328,438){$b$}
\put(190,450){\line(1,0){40}}
\put(240,450){\line(1,0){40}}
\put(290,450){\line(1,0){40}}
\put(260,450){\arc{40}{3.1416}{6.2832}}
\put(260,450){\arc{60}{3.1416}{6.2832}}
\put(225,415){$\la [ \rnode{1}{\mathbf{H}} ]_{c_{1} c_{2}} [ \rnode{2}{\mathbf{H}} ]_{c_{3} c_{4}} \ra \ncbar[linewidth=0.01,nodesep=2pt,arm=0.15,angle=90]{-}{1}{2}$}

\put(342,450){$+$}
\end{picture}

\begin{picture}(500,80)(0,335)
\put(102,370){$+$}

\put(120,370){\circle*{6}}
\put(118,358){$a$}
\put(160,370){\circle*{6}}
\put(157,358){$c_{1}$}
\put(170,370){\circle*{6}}
\put(167,358){$c_{2}$}
\put(210,370){\circle*{6}}
\put(207,358){$c_{3}$}
\put(220,370){\circle*{6}}
\put(217,358){$c_{4}$}
\put(260,370){\circle*{6}}
\put(257,358){$c_{5}$}
\put(270,370){\circle*{6}}
\put(267,358){$c_{6}$}
\put(310,370){\circle*{6}}
\put(307,358){$c_{7}$}
\put(320,370){\circle*{6}}
\put(317,358){$c_{8}$}
\put(360,370){\circle*{6}}
\put(358,358){$b$}
\put(120,370){\line(1,0){40}}
\put(170,370){\line(1,0){40}}
\put(220,370){\line(1,0){40}}
\put(270,370){\line(1,0){40}}
\put(320,370){\line(1,0){40}}
\put(190,370){\arc{40}{3.1416}{6.2832}}
\put(190,370){\arc{60}{3.1416}{6.2832}}
\put(290,370){\arc{40}{3.1416}{6.2832}}
\put(290,370){\arc{60}{3.1416}{6.2832}}
\put(175,335){$\la [ \rnode{1}{\mathbf{H}} ]_{c_{1} c_{2}} [ \rnode{2}{\mathbf{H}} ]_{c_{3} c_{4}} [ \rnode{3}{\mathbf{H}} ]_{c_{5} c_{6}} [ \rnode{4}{\mathbf{H}} ]_{c_{7} c_{8}} \ra \ncbar[linewidth=0.01,nodesep=2pt,arm=0.15,angle=90]{-}{1}{2} \ncbar[linewidth=0.01,nodesep=2pt,arm=0.15,angle=90]{-}{3}{4}$}

\put(372,370){$+$}
\end{picture}

\begin{picture}(500,130)(0,205)
\put(102,240){$+$}

\put(120,240){\circle*{6}}
\put(118,228){$a$}
\put(160,240){\circle*{6}}
\put(157,228){$c_{1}$}
\put(170,240){\circle*{6}}
\put(167,228){$c_{2}$}
\put(210,240){\circle*{6}}
\put(207,228){$c_{3}$}
\put(220,240){\circle*{6}}
\put(217,228){$c_{4}$}
\put(260,240){\circle*{6}}
\put(257,228){$c_{5}$}
\put(270,240){\circle*{6}}
\put(267,228){$c_{6}$}
\put(310,240){\circle*{6}}
\put(307,228){$c_{7}$}
\put(320,240){\circle*{6}}
\put(317,228){$c_{8}$}
\put(360,240){\circle*{6}}
\put(358,228){$b$}
\put(120,240){\line(1,0){40}}
\put(170,240){\line(1,0){40}}
\put(220,240){\line(1,0){40}}
\put(270,240){\line(1,0){40}}
\put(320,240){\line(1,0){40}}
\put(240,240){\arc{140}{3.1416}{6.2832}}
\put(240,240){\arc{160}{3.1416}{6.2832}}
\put(240,240){\arc{40}{3.1416}{6.2832}}
\put(240,240){\arc{60}{3.1416}{6.2832}}
\put(175,205){$\la [ \rnode{1}{\mathbf{H}} ]_{c_{1} c_{2}} [ \rnode{2}{\mathbf{H}} ]_{c_{3} c_{4}} [ \rnode{3}{\mathbf{H}} ]_{c_{5} c_{6}} [ \rnode{4}{\mathbf{H}} ]_{c_{7} c_{8}} \ra \ncbar[linewidth=0.01,nodesep=2pt,arm=0.25,angle=90]{-}{1}{4} \ncbar[linewidth=0.01,nodesep=2pt,arm=0.15,angle=90]{-}{2}{3}$}

\put(372,240){$+$}
\end{picture}

\begin{picture}(500,115)(0,90)
\put(102,140){$+$}

\put(120,140){\circle*{6}}
\put(118,128){$a$}
\put(160,140){\circle*{6}}
\put(157,128){$c_{1}$}
\put(170,140){\circle*{6}}
\put(167,128){$c_{2}$}
\put(210,140){\circle*{6}}
\put(207,128){$c_{3}$}
\put(220,140){\circle*{6}}
\put(217,128){$c_{4}$}
\put(260,140){\circle*{6}}
\put(257,128){$c_{5}$}
\put(270,140){\circle*{6}}
\put(267,128){$c_{6}$}
\put(310,140){\circle*{6}}
\put(307,128){$c_{7}$}
\put(320,140){\circle*{6}}
\put(317,128){$c_{8}$}
\put(360,140){\circle*{6}}
\put(358,128){$b$}
\put(120,140){\line(1,0){40}}
\put(170,140){\line(1,0){40}}
\put(220,140){\line(1,0){40}}
\put(270,140){\line(1,0){40}}
\put(320,140){\line(1,0){40}}
\put(210,140){\arc{100}{3.1416}{6.2832}}
\put(220,140){\arc{100}{3.1416}{6.2832}}
\put(260,140){\arc{100}{3.1416}{6.2832}}
\put(270,140){\arc{100}{3.1416}{6.2832}}
\put(175,105){$\la [ \rnode{1}{\mathbf{H}} ]_{c_{1} c_{2}} [ \rnode{2}{\mathbf{H}} ]_{c_{3} c_{4}} [ \rnode{3}{\mathbf{H}} ]_{c_{5} c_{6}} [ \rnode{4}{\mathbf{H}} ]_{c_{7} c_{8}} \ra \ncbar[linewidth=0.01,nodesep=2pt,arm=0.15,angle=90]{-}{1}{3} \ncbar[linewidth=0.01,nodesep=2pt,arm=0.25,angle=90]{-}{2}{4}$}
{\color{red}
\put(0,0){\path(130,180)(350,100)}
\put(0,0){\path(130,100)(350,180)}
\put(355,180){non--planar}}

\put(372,140){$+ \ldots$.}
\end{picture}
(There is summation over all the internal indices.) At the first order, we just have one horizontal line corresponding to \smash{$[ \mathbf{Z}^{- 1} ]_{a b}$}; at the second one, there are three \smash{$\mathbf{Z}^{- 1}$}'s (\ieNotAPPB three horizontal lines) and one propagator (\ieNotAPPB one double arc); while the third order contributes five \smash{$\mathbf{Z}^{- 1}$}'s (\ieNotAPPB five horizontal lines), and a $4$--point correlation function, which permits three possible contractions, giving rise to three ``rainbow'' diagrams, down to products of two propagators (\ieNotAPPB two double arcs), \emph{etc.}

Clearly, complete evaluation of \smash{$[ \mathbf{G}_{\mathbf{H}} ( z ) ]_{a b}$}, for any finite matrix size $N$, would require summing up the numerical values stemming from all such connected rainbow diagrams with external indices $a$ and $b$. This is where the thermodynamic limit of $N \to \infty$ comes into play --- as is well--known since the work of 't Hooft~\cite{tHooft74}, in this limit, only \emph{planar} graphs contribute to the Green's function, while all the non--planar ones are suppressed with the factor of \smash{$1 / N^{2 h}$}, where $h$ is the genus of the surface on which a given diagram can be drawn without crossing of lines. For instance, the last graph pictured above can only be drawn without a crossing on a torus (genus $1$), so it may be safely neglected at large $N$.

Now, all the planar diagrams can be summed up by exploiting a certain technique known from quantum field theory, which we will now sketch. First, one considers a subset of the rainbow diagrams, referred to as ``one--line--irreducible'' (1LI) ones, that cannot be split into two disconnected pieces by cutting a single horizontal line. Let \smash{$\mathbf{\Sigma}_{\mathbf{H}} ( z )$}, called the ``self--energy matrix'' (it is an $N \times N$ matrix), denote their generating function,

\begin{picture}(500,130)
\thicklines

\filltype{shade}
\put(50,35){\ellipse*{80}{40}}
\filltype{black}
\put(10,35){\circle*{6}}
\put(90,35){\circle*{6}}
\put(46,32){{\color{white}$\mathbf{\Sigma}$}}

\put(102,35){$=$}

\put(120,35){\circle*{6}}
\put(130,35){\circle*{6}}
\put(170,35){\circle*{6}}
\put(180,35){\circle*{6}}
\put(130,35){\line(1,0){40}}
\put(150,35){\arc{40}{3.1416}{6.2832}}
\put(150,35){\arc{60}{3.1416}{6.2832}}

\put(192,35){$+$}

\put(210,35){\circle*{6}}
\put(220,35){\circle*{6}}
\put(260,35){\circle*{6}}
\put(270,35){\circle*{6}}
\put(310,35){\circle*{6}}
\put(320,35){\circle*{6}}
\put(360,35){\circle*{6}}
\put(370,35){\circle*{6}}
\put(220,35){\line(1,0){40}}
\put(270,35){\line(1,0){40}}
\put(320,35){\line(1,0){40}}
\put(290,35){\arc{40}{3.1416}{6.2832}}
\put(290,35){\arc{60}{3.1416}{6.2832}}
\put(290,35){\arc{140}{3.1416}{6.2832}}
\put(290,35){\arc{160}{3.1416}{6.2832}}

\put(382,35){$+ \ldots$.}
\end{picture}

One now perceives that any planar rainbow diagram can be constructed by putting together horizontal lines and 1LI graphs, \ieNotAPPB a relationship between the full and 1LI generating functions pictorially reads

\begin{picture}(500,55)(0,15)
\thicklines

\put(10,35){\circle*{6}}
\put(90,35){\circle*{6}}
\put(10,35){\line(1,0){20}}
\put(90,35){\line(-1,0){20}}
\filltype{shade}
\put(50,35){\circle*{40}}
\filltype{black}
\put(46,32){{\color{white}$\mathbf{G}$}}

\put(102,35){$=$}

\put(120,35){\circle*{6}}
\put(160,35){\circle*{6}}
\put(120,35){\line(1,0){40}}

\put(172,35){$+$}

\filltype{shade}
\put(270,35){\ellipse*{80}{40}}
\filltype{black}
\put(190,35){\circle*{6}}
\put(230,35){\circle*{6}}
\put(310,35){\circle*{6}}
\put(350,35){\circle*{6}}
\put(190,35){\line(1,0){40}}
\put(310,35){\line(1,0){40}}
\put(266,32){{\color{white}$\mathbf{\Sigma}$}}

\put(362,35){$+$}
\end{picture}

\begin{picture}(500,70)
\thicklines

\put(102,35){$+$}

\filltype{shade}
\put(200,35){\ellipse*{80}{40}}
\put(320,35){\ellipse*{80}{40}}
\filltype{black}
\put(120,35){\circle*{6}}
\put(160,35){\circle*{6}}
\put(240,35){\circle*{6}}
\put(280,35){\circle*{6}}
\put(360,35){\circle*{6}}
\put(400,35){\circle*{6}}
\put(120,35){\line(1,0){40}}
\put(240,35){\line(1,0){40}}
\put(360,35){\line(1,0){40}}
\put(196,32){{\color{white}$\mathbf{\Sigma}$}}
\put(316,32){{\color{white}$\mathbf{\Sigma}$}}

\put(412,35){$+ \ldots$,}
\end{picture}
\ieNotAPPB
\begin{equation}\label{eq:GUEDysonSchwinger1}
\mathbf{G}_{\mathbf{H}} ( z ) = \mathbf{Z}^{- 1} + \mathbf{Z}^{- 1} \mathbf{\Sigma}_{\mathbf{H}} ( z ) \mathbf{Z}^{- 1} + \mathbf{Z}^{- 1} \mathbf{\Sigma}_{\mathbf{H}} ( z ) \mathbf{Z}^{- 1} \mathbf{\Sigma}_{\mathbf{H}} ( z ) \mathbf{Z}^{- 1} + \ldots = \left( \mathbf{Z} - \mathbf{\Sigma}_{\mathbf{H}} ( z ) \right)^{- 1} .
\end{equation}
This is the first Dyson--Schwinger's equation. Remark that it looks very similar to (\ref{eq:GreensFunctionMatrixExpansion}), (\ref{eq:GreensFunctionMatrixDefinition}), except for the now--absent averaging; thus, we may say that the self--energy matrix represents an effective matrix $\mathbf{H}$ which allows one to remove the averaging operation from (\ref{eq:GreensFunctionMatrixExpansion}), (\ref{eq:GreensFunctionMatrixDefinition}).

Another observation is that if one takes an arbitrary planar rainbow diagram and adds an external double arc to it, one obtains a 1LI graph; and conversely, any 1LI graph has a form of a double arc encircling some diagram. Hence, the generating functions in question satisfy

\begin{picture}(500,100)
\thicklines

\filltype{shade}
\put(50,35){\ellipse*{80}{40}}
\filltype{black}
\put(10,35){\circle*{6}}
\put(8,23){$a$}
\put(90,35){\circle*{6}}
\put(88,23){$b$}
\put(46,32){{\color{white}$\mathbf{\Sigma}$}}

\put(102,35){$=$}

\put(120,35){\circle*{6}}
\put(118,23){$a$}
\put(130,35){\circle*{6}}
\put(127,23){$c_{1}$}
\put(210,35){\circle*{6}}
\put(207,23){$c_{2}$}
\put(220,35){\circle*{6}}
\put(218,23){$b$}
\put(130,35){\line(1,0){20}}
\put(210,35){\line(-1,0){20}}
\filltype{shade}
\put(170,35){\circle*{40}}
\filltype{black}
\put(166,32){{\color{white}$\mathbf{G}$}}
\put(170,35){\arc{80}{3.1416}{6.2832}}
\put(170,35){\arc{100}{3.1416}{6.2832}}
\end{picture}
which is
$$
[ \mathbf{\Sigma}_{\mathbf{H}} ( z ) ]_{a b} = \sum_{c_{1} , c_{2} = 1}^{N} [ \mathbf{G}_{\mathbf{H}} ( z ) ]_{c_{1} c_{2}} \la [ \mathbf{H} ]_{a c_{1}} [ \mathbf{H} ]_{c_{2} b} \ra = \sigma^{2} \left( \frac{1}{N} \Tr \mathbf{G}_{\mathbf{H}} ( z ) \right) \delta_{a b} ,
$$
\begin{equation}\label{eq:GUEDysonSchwinger2}
\textrm{\ieNotAPPB} \quad \mathbf{\Sigma}_{\mathbf{H}} ( z ) = \sigma^{2} G_{\mathbf{H}} ( z ) \Id_{N} .
\end{equation}
This is the second Dyson--Schwinger's equation. Since it reveals that the self--energy matrix is proportional to the unit matrix, one may take the normalized trace of (\ref{eq:GUEDysonSchwinger2}) and rewrite it as
\begin{equation}\label{eq:GUEDysonSchwinger2Tr}
\Sigma_{\mathbf{H}} ( z ) = \sigma^{2} G_{\mathbf{H}} ( z ) ,
\end{equation}
where, analogously to (\ref{eq:GreensFunctionDefinition}), the ``self--energy,''
\begin{equation}\label{eq:SelfEnergyDefinition}
\Sigma_{\mathbf{H}} ( z ) \equiv \frac{1}{N} \Tr \mathbf{\Sigma}_{\mathbf{H}} ( z ) .
\end{equation}

Finally, substituting \smash{$\mathbf{\Sigma}_{\mathbf{H}} ( z )$} from the second Dyson--Schwinger's equation (\ref{eq:GUEDysonSchwinger2}) into the first one (\ref{eq:GUEDysonSchwinger1}), and applying the normalized trace to both sides --- yields a quadratic equation for the Green's function of the GUE ensemble (\ref{eq:GUEMeasure}),
\begin{equation}\label{eq:GreensFunctionForGUE}
G_{\mathbf{H}} ( z ) \left( z - \sigma^{2} G_{\mathbf{H}} ( z ) \right) = 1 , \quad \textrm{\ieNotAPPB} \quad G_{\mathbf{H}} ( z ) = \frac{1}{2 \sigma^{2}} \left( z - \sqrt{z - 2 \sigma} \sqrt{z + 2 \sigma} \right) ,
\end{equation}
where we have selected the correct solution out of the two by referring to the condition \smash{$G_{\mathbf{H}} ( z ) \sim 1 / z$}, for $z \to \infty$ (equivalent to the normalization of the density, \smash{$\int_{\mathrm{cuts}} \dd \lambda \rho_{\mathbf{H}} ( \lambda ) = 1$}, see (\ref{eq:GreensFunctionFromMeanSpectralDensity})); also, the square roots are in the principal branches. The average eigenvalue density is found from this Green's function using (\ref{eq:MeanSpectralDensityFromGreensFunction}), and is the famous ``Wigner's semi--circle distribution,''
\begin{equation}\label{eq:WignersSemiCircleDistribution}
\rho_{\mathbf{H}} ( \lambda ) = \left\{ \begin{array}{ll} \frac{1}{2 \pi \sigma^{2}} \sqrt{4 \sigma^{2} - \lambda^{2}} , & \quad \textrm{for} \quad \lambda \in [ - 2 \sigma , 2 \sigma ] , \\ 0 , & \quad \textrm{for} \quad \lambda \in \mathbb{R} \setminus [ - 2 \sigma , 2 \sigma ] . \end{array} \right.
\end{equation}

On this pedagogical example, we have shown how planar diagrammatics and Dyson--Schwinger's equations can be used to solve Hermitian matrix models in the thermodynamic limit. Let us now see whether non--Hermitian ensembles could be appropriated along similar lines.


\subsection{\ Non--Hermitian Random Matrices}
\label{ss:NonHermitianRandomMatrices}


\subsubsection{\ The Non--Holomorphic Green's Function and $M$--Transform}
\label{sss:TheNonHolomorphicGreensFunctionAndMTransform}

The chief difference between Hermitian and non--Hermitian random matrices is that the eigenvalues \smash{$\lambda_{1} , \lambda_{2} , \ldots , \lambda_{N}$} of the latter (call it $\mathbf{X}$) are \emph{complex} --- in the thermodynamic limit of $N \to \infty$ merging into some two--dimensional ``domains'' $\mathcal{D}$ on the complex plane --- with their average distribution given by the complex Dirac's delta,
\begin{equation}\label{eq:NonHermitianMeanSpectralDensityDefinition}
\rho_{\mathbf{X}} ( \lambda , \overline{\lambda} ) \equiv \frac{1}{N} \sum_{a = 1}^{N} \la \delta^{( 2 )} \left( \lambda - \lambda_{a} , \overline{\lambda} - \overline{\lambda_{a}} \right) \ra .
\end{equation}
Thus, the concepts developed in \S\ref{sss:TheGreensFunctionAndMTransform} and relying on the reality of the eigenvalues, the Green's function and $M$--transform, lose their meaning.

The bottom line is that the complex Dirac's delta has distinct properties from its real counterpart. For instance, its useful representation will now be
\begin{equation}\label{eq:ComplexDiracsDelta}
\delta^{( 2 )} ( \lambda , \overline{\lambda} ) = \frac{1}{\pi} \lim_{\epsilon \to 0} \frac{\epsilon^{2}}{\left( | \lambda |^{2} + \epsilon^{2} \right)^{2}} = \frac{1}{\pi} \frac{\partial}{\partial \overline{\lambda}} \lim_{\epsilon \to 0} \frac{\overline{\lambda}}{| \lambda |^{2} + \epsilon^{2}} ,
\end{equation}
as contrasted with (\ref{eq:RealDiracsDelta}). Mimicking this form, one is led to define the ``non--holomorphic Green's function'' as
$$
G_{\mathbf{X}} ( z , \overline{z} ) \equiv \lim_{\epsilon \to 0} \lim_{N \to \infty} \frac{1}{N} \sum_{a = 1}^{N} \la \frac{\overline{z} - \overline{\lambda_{a}}}{\left| z - \lambda_{a} \right|^{2} + \epsilon^{2}} \ra =
$$
\begin{equation}\label{eq:NonHolomorphicGreensFunctionDefinition}
= \lim_{\epsilon \to 0} \lim_{N \to \infty} \frac{1}{N} \Tr \la \frac{\overline{z} \Id_{N} - \mathbf{X}^{\dagger}}{\left( z \Id_{N} - \mathbf{X} \right) \left( \overline{z} \Id_{N} - \mathbf{X}^{\dagger} \right) + \epsilon^{2} \Id_{N}} \ra
\end{equation}
(we have ambiguously written here the matrix inversion as a fraction --- we assume henceforth \smash{$\mathbf{A} / \mathbf{B} \equiv \mathbf{A} \mathbf{B}^{- 1}$}, \ieNotAPPB multiplication by the inverse matrix from the right), since then the density (\ref{eq:NonHermitianMeanSpectralDensityDefinition}) is just proportional to its derivative w.r.t. \smash{$\overline{z}$},
\begin{equation}\label{eq:MeanSpectralDensityFromNonHolomorphicGreensFunction}
\rho_{\mathbf{X}} ( z , \overline{z} ) = \frac{1}{\pi} \frac{\partial}{\partial \overline{z}} G_{\mathbf{X}} ( z , \overline{z} ) , \quad \textrm{for} \quad z \in \mathcal{D} .
\end{equation}
Notice the order of the limits in (\ref{eq:NonHolomorphicGreensFunctionDefinition}) --- for finite $N$, one could immediately set the regulator $\epsilon$ to zero everywhere except a finite set of points \smash{$\lambda_{a}$}, and the non--holomorphic Green's function would reduce to the usual one (\ref{eq:GreensFunctionDefinition}); only for infinite $N$ and inside the eigenvalues' domains $\mathcal{D}$, the limit $\epsilon \to 0$ yields a non--trivial object, carrying information about the mean eigenvalue density. Outside $\mathcal{D}$, one is always left with the usual Green's function (\ref{eq:GreensFunctionDefinition}),
\begin{equation}\label{eq:NonHolomorphicGreensFunctionOutsideD}
G_{\mathbf{X}} ( z , \overline{z} ) = G_{\mathbf{X}} ( z ) , \quad \textrm{for} \quad z \notin \mathcal{D} ,
\end{equation}
which however does not tell this time anything about the eigenvalues. Actually, this formula holds on the boundary of $\mathcal{D}$ as well,
\begin{equation}\label{eq:NonHolomorphicGreensFunctionOnTheBoundaryOfD}
G_{\mathbf{X}} ( z , \overline{z} ) = G_{\mathbf{X}} ( z ) , \quad \textrm{for} \quad z \in \partial \mathcal{D} ,
\end{equation}
which we will see to be precisely the equation fulfilled by the coordinates of the boundary of $\mathcal{D}$.

Similarly as in the Hermitian case, one finds it useful to consider a related quantity, named the ``non--holomorphic $M$--transform,'' and defined analogously to (\ref{eq:MTransformDefinition}),
\begin{equation}\label{eq:NonHolomorphicMTransformDefinition}
M_{\mathbf{X}} ( z , \overline{z} ) \equiv z G_{\mathbf{X}} ( z , \overline{z} ) - 1 .
\end{equation}
Again, outside of $\mathcal{D}$ it simplifies to the usual $M$--transform \smash{$M_{\mathbf{X}} ( z )$}, which is however obviously incapable of capturing the distribution of the eigenvalues of $\mathbf{X}$.

Important as it is, an essential hindrance in an effective usage of the non--holomorphic Green's function (\ref{eq:NonHolomorphicGreensFunctionDefinition}) is the quadratic (in $\mathbf{X}$) structure of its denominator. As a response to that, the following \emph{linearization} trick has been proposed: Introduce a $2 N \times 2 N$ matrix
\begin{equation}\label{eq:MatrixValuedGreensFunctionMatrixDefinition1}
\mathbf{G}^{\DD}_{\mathbf{X}} ( z , \overline{z} ) \equiv \lim_{\epsilon \to 0} \lim_{N \to \infty} \la \left( \mathbf{Z}^{\DD}_{\epsilon} - \mathbf{X}^{\DD} \right)^{- 1} \ra ,
\end{equation}
where
\begin{equation}\label{eq:MatrixValuedGreensFunctionMatrixDefinition2}
\mathbf{Z}^{\DD}_{\epsilon} \equiv \left( \begin{array}{cc} z \Id_{N} & \ii \epsilon \Id_{N} \\ \ii \epsilon \Id_{N} & \overline{z} \Id_{N} \end{array} \right) , \quad \mathbf{X}^{\DD} \equiv \left( \begin{array}{cc} \mathbf{X} & \Zero_{N} \\ \Zero_{N} & \mathbf{X}^{\dagger} \end{array} \right) .
\end{equation}
The superscript ``$\DD$'' comes from ``\underline{d}uplication.'' Considering the four $N \times N$ blocks of this matrix, distinguished by the superscripts $z z$, \smash{$z \overline{z}$}, \smash{$\overline{z} z$} and \smash{$\overline{z} \overline{z}$} (and omitting the symbols of the functional dependence on $z , \overline{z}$, which we will often do in similar expressions),
\begin{equation}\label{eq:MatrixValuedGreensFunctionMatrixBlocksDefinition}
\mathbf{G}^{\DD}_{\mathbf{X}} \equiv \left( \begin{array}{cc} \mathbf{G}^{z z}_{\mathbf{X}} & \mathbf{G}^{z \overline{z}}_{\mathbf{X}} \\ \mathbf{G}^{\overline{z} z}_{\mathbf{X}} & \mathbf{G}^{\overline{z} \overline{z}}_{\mathbf{X}} \end{array} \right) ,
\end{equation}
as well as its normalized ``block--trace,''
\begin{equation}\label{eq:BlockTraceDefinition}
\bTr \left( \begin{array}{cc} \mathbf{A} & \mathbf{B} \\ \mathbf{C} & \mathbf{D} \end{array} \right) \equiv \left( \begin{array}{cc} \Tr \mathbf{A} & \Tr \mathbf{B} \\ \Tr \mathbf{C} & \Tr \mathbf{D} \end{array} \right) ,
\end{equation}
\ieNotAPPB a $2 \times 2$ matrix, referred to as the ``matrix--valued Green's function,'' consisting of the normalized traces of the four blocks,
\begin{equation}\label{eq:MatrixValuedGreensFunctionDefinition}
\mathcal{G}_{\mathbf{X}} \equiv \frac{1}{N} \bTr \mathbf{G}^{\DD}_{\mathbf{X}} = \left( \begin{array}{cc} \frac{1}{N} \Tr \mathbf{G}^{z z}_{\mathbf{X}} & \frac{1}{N} \Tr \mathbf{G}^{z \overline{z}}_{\mathbf{X}} \\ \frac{1}{N} \Tr \mathbf{G}^{\overline{z} z}_{\mathbf{X}} & \frac{1}{N} \Tr \mathbf{G}^{\overline{z} \overline{z}}_{\mathbf{X}} \end{array} \right) \equiv \left( \begin{array}{cc} \mathcal{G}^{z z}_{\mathbf{X}} & \mathcal{G}^{z \overline{z}}_{\mathbf{X}} \\ \mathcal{G}^{\overline{z} z}_{\mathbf{X}} & \mathcal{G}^{\overline{z} \overline{z}}_{\mathbf{X}} \end{array} \right) ,
\end{equation}
one recognizes that the upper left corner of the matrix--valued Green's function is exactly equal to the non--holomorphic Green's function (\ref{eq:NonHolomorphicGreensFunctionDefinition}),
\begin{equation}\label{eq:MatrixValuedGreensFunctionBlockZZ}
\mathcal{G}^{z z}_{\mathbf{X}} ( z , \overline{z} ) = G_{\mathbf{X}} ( z , \overline{z} ) .
\end{equation}
In this way, the relevant non--holomorphic Green's function has been encoded as a part of another object, having linear structure in $\mathbf{X}$, and looking very similar to the usual Green's function (compare (\ref{eq:MatrixValuedGreensFunctionMatrixDefinition1}) with (\ref{eq:GreensFunctionMatrixDefinition})), the price to pay for that being ``duplication'' of the matrices. This will enable us to use techniques such as planar diagrammatics to compute the matrix--valued Green's function of a non--Hermitian ensemble $\mathbf{X}$ in a manner parallel to how one does it in the Hermitian sector (\S\ref{sss:TheGreensFunctionForTheGUEFromPlanarDiagrams}); see below (\S\ref{sss:TheNonHolomorphicGreensFunctionForTheGirkoGinibreEnsembleFromPlanarDiagrams}).

Let us close with the following remark: We have seen that the $z z$--component of the matrix--valued Green's function is equal to the non--holomorphic Green's function (\ref{eq:MatrixValuedGreensFunctionBlockZZ}), and thus, inside $\mathcal{D}$ reproduces the mean eigenvalue density (\ref{eq:MeanSpectralDensityFromNonHolomorphicGreensFunction}), while outside $\mathcal{D}$ reduces to the usual Green's function (\ref{eq:NonHolomorphicGreensFunctionOutsideD}). The \smash{$\overline{z} \overline{z}$}--component is just its complex conjugate, \smash{$\mathcal{G}^{\overline{z} \overline{z}}_{\mathbf{X}} = \overline{\mathcal{G}^{z z}_{\mathbf{X}}}$}, so it carries no additional information. One might wonder about the off--diagonal elements,
\begin{equation}\label{eq:MatrixValuedGreensFunctionBlockZZBar}
\mathcal{G}^{z \overline{z}}_{\mathbf{X}} ( z , \overline{z} ) = \lim_{\epsilon \to 0} \lim_{N \to \infty} \frac{1}{N} \Tr \la \frac{- \ii \epsilon}{\left( \overline{z} \Id_{N} - \mathbf{X}^{\dagger} \right) \left( z \Id_{N} - \mathbf{X} \right) + \epsilon^{2} \Id_{N}} \ra ,
\end{equation}
\begin{equation}\label{eq:MatrixValuedGreensFunctionBlockZBarZ}
\mathcal{G}^{\overline{z} z}_{\mathbf{X}} ( z , \overline{z} ) = \lim_{\epsilon \to 0} \lim_{N \to \infty} \frac{1}{N} \Tr \la \frac{- \ii \epsilon}{\left( z \Id_{N} - \mathbf{X} \right) \left( \overline{z} \Id_{N} - \mathbf{X}^{\dagger} \right) + \epsilon^{2} \Id_{N}} \ra .
\end{equation}
Since they are equal to each other and purely imaginary, \smash{$\ii \mathcal{G}^{z \overline{z}}_{\mathbf{X}} ( z , \overline{z} ) = \ii \mathcal{G}^{\overline{z} z}_{\mathbf{X}} ( z , \overline{z} ) \in \mathbb{R}$}, it is convenient to consider their negated product, which is a non--negative real number,
\begin{equation}\label{eq:MatrixValuedGreensFunctionMinusProductOfOffDiagonalElements}
\mathcal{C}_{\mathbf{X}} ( z , \overline{z} ) \equiv - \mathcal{G}^{z \overline{z}}_{\mathbf{X}} ( z , \overline{z} ) \mathcal{G}^{\overline{z} z}_{\mathbf{X}} ( z , \overline{z} ) \in \mathbb{R}_{+} \cup \{ 0 \} .
\end{equation}
Clearly, it vanishes outside of the domains of the eigenvalues $\mathcal{D}$, as the regulator may be safely zeroed there,
\begin{equation}\label{eq:MatrixValuedGreensFunctionBlocksZZBarAndZBarZOutsideD}
\mathcal{C}_{\mathbf{X}} ( z , \overline{z} ) = 0 , \quad \textrm{for} \quad z \notin \mathcal{D} ,
\end{equation}
whereas inside $\mathcal{D}$, it acquires non--trivial (strictly positive) values. Consequently, it contains all the information about the boundary of $\mathcal{D}$ --- if one finds \smash{$\mathcal{C}_{\mathbf{X}} ( z , \overline{z} )$} inside $\mathcal{D}$, and subsequently sets it to $0$ (\ref{eq:MatrixValuedGreensFunctionBlocksZZBarAndZBarZOutsideD}), one obtains an equation obeyed by the coordinates $( x , y )$, where $z \equiv x + \ii y$, of $\partial \mathcal{D}$. (Note that this is equivalent to (\ref{eq:NonHolomorphicGreensFunctionOnTheBoundaryOfD}).) This being our primary usage of \smash{$\mathcal{C}_{\mathbf{X}} ( z , \overline{z} )$}, let us nevertheless mention that it has yet another content --- it describes certain statistical features of the left and right eigenvectors of $\mathbf{X}$~\cite{ChalkerMehlig1998}.


\subsubsection{\ The Non--Holomorphic Green's Function for the Girko--Ginibre Ensemble from Planar Diagrams}
\label{sss:TheNonHolomorphicGreensFunctionForTheGirkoGinibreEnsembleFromPlanarDiagrams}

Since the matrix--valued and usual Green's functions have analogous forms, the difference being the duplicated structure in the latter case, planar diagrammatics and the Dyson--Schwinger's equations will work in a similar way. Instead of giving a general introduction, let us present the method on a simplest example --- the Girko--Ginibre matrix model $\mathbf{A}$ (with standard deviation $\sigma$), defined in (\ref{eq:SquareGGMeasure}).

The propagators are immediately read off from this measure (\ref{eq:SquareGGMeasure}),
$$
\la [ \mathbf{A} ]_{a b} \left[ \mathbf{A}^{\dagger} \right]_{c d} \ra = \la \left[ \mathbf{A}^{\dagger} \right]_{a b} [ \mathbf{A} ]_{c d} \ra = \frac{\sigma^{2}}{N} \delta_{a d} \delta_{b c} ,
$$
\begin{equation}\label{eq:GGPropagators}
\la [ \mathbf{A} ]_{a b} [ \mathbf{A} ]_{c d} \ra = \la \left[ \mathbf{A}^{\dagger} \right]_{a b} \left[ \mathbf{A}^{\dagger} \right]_{c d} \ra = 0 .
\end{equation}
However, in order to employ the diagrammatic tools, one should know the propagators of the duplicated matrix \smash{$\mathbf{A}^{\DD}$} (\ref{eq:MatrixValuedGreensFunctionMatrixDefinition1}). Denoting the indices in its four $N \times N$ blocks by $a b$, \smash{$a \overline{b}$}, \smash{$\overline{a} b$}, \smash{$\overline{a} \overline{b}$}, respectively, mimicking the labeling of $z z$, \smash{$z \overline{z}$}, \smash{$\overline{z} z$}, \smash{$\overline{z} \overline{z}$} in (\ref{eq:MatrixValuedGreensFunctionMatrixBlocksDefinition}), one sees that the only non--zero propagators read
\begin{equation}\label{eq:GGDPropagators}
\la \left[ \mathbf{A}^{\DD} \right]_{a b} \left[ \mathbf{A}^{\DD} \right]_{\overline{c} \overline{d}} \ra = \frac{\sigma^{2}}{N} \delta_{a \overline{d}} \delta_{b \overline{c}} , \quad \la \left[ \mathbf{A}^{\DD} \right]_{\overline{a} \overline{b}} \left[ \mathbf{A}^{\DD} \right]_{c d} \ra = \frac{\sigma^{2}}{N} \delta_{\overline{a} d} \delta_{\overline{b} c} .
\end{equation}

This is all there is required in order to write down the Dyson--Schwinger's equations. The self--energy matrix will be duplicated,
\begin{equation}\label{eq:DuplicatedSelfEnergyMatrixDefinition}
\mathbf{\Sigma}^{\DD} \equiv \left( \begin{array}{cc} \mathbf{\Sigma}^{z z} & \mathbf{\Sigma}^{z \overline{z}} \\ \mathbf{\Sigma}^{\overline{z} z} & \mathbf{\Sigma}^{\overline{z} \overline{z}} \end{array} \right)
\end{equation}
(for simplicity, we remove the subscript ``$\mathbf{A}$'' in this paragraph). The first Dyson--Schwinger's equation will clearly have an identical form to (\ref{eq:GUEDysonSchwinger1}),
\begin{equation}\label{eq:GGDysonSchwinger1}
\mathbf{G}^{\DD} = \left( \mathbf{Z}^{\DD} - \mathbf{\Sigma}^{\DD} \right)^{- 1} ,
\end{equation}
where \smash{$\mathbf{Z}^{\DD} \equiv \mathbf{Z}^{\DD}_{\epsilon = 0}$} (\ref{eq:MatrixValuedGreensFunctionMatrixDefinition2}) --- the regulator $\epsilon$ may henceforth be set to zero; we will see that the Dyson--Schwinger's equations by themselves take care of properly regulating the result inside the mean eigenvalues' domains $\mathcal{D}$. The second Dyson--Schwinger's equation will also pictorially look like before (see the diagrammatic equation above (\ref{eq:GUEDysonSchwinger2})), but the present structure of the propagators (\ref{eq:GGDPropagators}) implies that the four blocks of that formula now read
\begin{equation}\label{eq:GGDysonSchwinger2BlockZZBar}
\left[ \mathbf{\Sigma}^{\DD} \right]_{a \overline{b}} = \sum_{c_{1} , \overline{c_{2}} = 1}^{N} \left[ \mathbf{G}^{\DD} \right]_{c_{1} \overline{c_{2}}} \la \left[ \mathbf{A}^{\DD} \right]_{a c_{1}} \left[ \mathbf{A}^{\DD} \right]_{\overline{c_{2}} \overline{b}} \ra = \sigma^{2} \left( \frac{1}{N} \Tr \mathbf{G}^{z \overline{z}} \right) \delta_{a \overline{b}} = \sigma^{2} \mathcal{G}^{z \overline{z}} \delta_{a \overline{b}} ,
\end{equation}
\begin{equation}\label{eq:GGDysonSchwinger2BlockZBarZ}
\left[ \mathbf{\Sigma}^{\DD} \right]_{\overline{a} b} = \sigma^{2} \mathcal{G}^{\overline{z} z} \delta_{\overline{a} b} ,
\end{equation}
\begin{equation}\label{eq:GGDysonSchwinger2BlocksZZAndZBarZBar}
\left[ \mathbf{\Sigma}^{\DD} \right]_{a b} = \left[ \mathbf{\Sigma}^{\DD} \right]_{\overline{a} \overline{b}} = 0 ,
\end{equation}
\ieNotAPPB altogether,
\begin{equation}\label{eq:GGDysonSchwinger2}
\mathbf{\Sigma}^{\DD} = \left( \begin{array}{cc} \Zero_{N} & \sigma^{2} \mathcal{G}^{z \overline{z}} \Id_{N} \\ \sigma^{2} \mathcal{G}^{\overline{z} z} \Id_{N} & \Zero_{N} \end{array} \right) .
\end{equation}

Substituting the expression (\ref{eq:GGDysonSchwinger2}) for the duplicated self--energy matrix into (\ref{eq:GGDysonSchwinger1}), and taking the normalized block--trace (\ref{eq:BlockTraceDefinition}) of both sides, we arrive at the $2 \times 2$ matrix equation for the elements of the matrix--valued Green's function of the Girko--Ginibre ensemble (\ref{eq:SquareGGMeasure}),
\begin{equation}\label{eq:MatrixValuedGreensFunctionForGGMatrixEquation}
\left( \begin{array}{cc} \mathcal{G}^{z z} & \mathcal{G}^{z \overline{z}} \\ \mathcal{G}^{\overline{z} z} & \mathcal{G}^{\overline{z} \overline{z}} \end{array} \right) = \left( \begin{array}{cc} z & - \sigma^{2} \mathcal{G}^{z \overline{z}} \\ - \sigma^{2} \mathcal{G}^{\overline{z} z} & \overline{z} \end{array} \right)^{- 1} = \frac{1}{| z |^{2} - \sigma^{4} \mathcal{G}^{z \overline{z}} \mathcal{G}^{\overline{z} z}} \left( \begin{array}{cc} \overline{z} & \sigma^{2} \mathcal{G}^{z \overline{z}} \\ \sigma^{2} \mathcal{G}^{\overline{z} z} & z \end{array} \right) ,
\end{equation}
where in the last step we have explicitly inverted the matrix. Let us first look at the negated product of the two off--diagonal equations of (\ref{eq:MatrixValuedGreensFunctionForGGMatrixEquation}), which yields an equation for \smash{$\mathcal{C} \equiv \mathcal{C}_{\mathbf{A}} ( z , \overline{z} )$} (\ref{eq:MatrixValuedGreensFunctionMinusProductOfOffDiagonalElements}),
\begin{equation}\label{eq:MatrixValuedGreensFunctionForGGOffDiagonalEquation}
\mathcal{C} = \frac{\sigma^{4} \mathcal{C}}{\left( | z |^{2} + \sigma^{4} \mathcal{C} \right)^{2}} .
\end{equation}
It has one solution $\mathcal{C} = 0$, corresponding to the outside of $\mathcal{D}$, and called the ``holomorphic solution,'' as well as a non--trivial one, describing the inside of $\mathcal{D}$, named the ``non--holomorphic solution,'' and given by
\begin{equation}\label{eq:MatrixValuedGreensFunctionForGGOffDiagonalEquationSolution}
\mathcal{C}_{\mathbf{A}} ( z , \overline{z} ) = \frac{1}{\sigma^{2}} \left( 1 - \frac{| z |^{2}}{\sigma^{2}} \right) ,
\end{equation}
where we have picked the positive root out of the two. As mentioned (see the discussion below (\ref{eq:MatrixValuedGreensFunctionBlocksZZBarAndZBarZOutsideD})), setting this quantity to zero yields the equation of the borderline of $\mathcal{D}$ --- since the holomorphic and non--holomorphic solutions must coincide on $\partial \mathcal{D}$,
\begin{equation}\label{eq:GGBoundary}
| z | = \sigma ,
\end{equation}
\ieNotAPPB the average eigenvalues of the Girko--Ginibre model fill the centered circle of radius $\sigma$. Let us now proceed to the one diagonal equation (stemming from the upper left corner) of (\ref{eq:MatrixValuedGreensFunctionForGGMatrixEquation}),
\begin{equation}\label{eq:MatrixValuedGreensFunctionForGGDiagonalEquationSolution}
\mathcal{G}^{z z} = \frac{\overline{z}}{| z |^{2} + \sigma^{4} \mathcal{C}} = \left\{ \begin{array}{ll} \frac{\overline{z}}{\sigma^{2}} , & \quad \textrm{for} \quad | z | \leq \sigma , \\ \frac{1}{z} , & \quad \textrm{for} \quad | z | > \sigma . \end{array} \right.
\end{equation}
We explicitly see that outside of the eigenvalues' disk, it is trivial, equal to the usual Green's function (the ``holomorphic solution''), which for the Girko--Ginibre ensemble is just $1 / z$. However, inside the disk, it provides the non--holomorphic Green's function (\ref{eq:MatrixValuedGreensFunctionBlockZZ}) of the model,
\begin{equation}\label{eq:MatrixValuedGreensFunctionForGGDiagonalEquationSolution}
G_{\mathbf{A}} ( z , \overline{z} ) = \frac{\overline{z}}{\sigma^{2}} ,
\end{equation}
and subsequently (\ref{eq:MeanSpectralDensityFromNonHolomorphicGreensFunction}), the mean eigenvalue density (\ref{eq:GGMeanSpectralDensity}).

The Girko--Ginibre model is therefore solved in the thermodynamic limit by means of planar diagrammatics on a ``duplicated'' level, outlining a pattern to follow when dealing with more complicated non--Hermitian random matrices.


\subsubsection{\ Non--Hermitian Planar Diagrammatics for Products of Random Matrices}
\label{sss:NonHermitianPlanarDiagrammaticsForProductsOfRandomMatrices}

As explained in~\S\ref{sss:TheLinearizationOfTheProduct}, if one wants to apply the method of planar diagrams to a model being a product of random matrices, the linearization (\ref{eq:EIG01}) is necessary. In this paragraph, we will prove the main result of this procedure, (\ref{eq:EIG07a}).

Since the $L$--th power of (\ref{eq:EIG01}) is an $L \times L$ block--diagonal matrix (total size of course \smash{$N_{\tot} \times N_{\tot}$}), with the entries being all the cyclic permutations of the product (\ref{eq:ProductDefinition}),
\begin{equation}\label{eq:EIG02}
\widetilde{\mathbf{P}}^{L} = \left( \begin{array}{cccc} \mathbf{A}_{1} \mathbf{A}_{2} \ldots \mathbf{A}_{L - 1} \mathbf{A}_{L} & \Zero & \ldots & \Zero \\ \Zero & \mathbf{A}_{2} \mathbf{A}_{3} \ldots \mathbf{A}_{L} \mathbf{A}_{1} & \ldots & \Zero \\ \vdots & \vdots & \ddots & \vdots \\ \Zero & \Zero & \ldots & \mathbf{A}_{L} \mathbf{A}_{1} \ldots \mathbf{A}_{L - 2} \mathbf{A}_{L - 1} \end{array} \right) ,
\end{equation}
and because all these permutations have identical non--zero eigenvalues, plus a number (distinct for all these matrices) of zero modes --- therefore, \smash{$\widetilde{\mathbf{P}}^{L}$} has the same eigenvalues as $\mathbf{P}$, name them \smash{$\lambda_{1} , \lambda_{2} , \ldots , \lambda_{N_{1}}$}, each one $L$--fold degenerate (so in total \smash{$L N_{1}$}), plus \smash{$( N_{\tot} - L N_{1} )$} additional zero modes (some of the \smash{$\lambda_{a}$}'s may be zero as well). This implies that the content of the \smash{$N_{\tot}$} eigenvalues of \smash{$\widetilde{\mathbf{P}}$} is the following: \smash{$L N_{1}$} eigenvalues \smash{$\sqrt[L]{\lambda_{1}} , \sqrt[L]{\lambda_{2}} , \ldots , \sqrt[L]{\lambda_{N_{1}}}$} (each one with $L$--fold degeneracy), where we keep in mind that the $L$--th root is an $L$--valued function, plus \smash{$( N_{\tot} - L N_{1} )$} additional zero modes. Thus, we may translate this description into a mathematical expression (\ref{eq:NonHermitianMeanSpectralDensityDefinition}),
$$
\rho_{\widetilde{\mathbf{P}}} ( w , \overline{w} ) = \frac{1}{N_{\tot}} \left( L \sum_{a = 1}^{N_{1}} \la \delta^{( 2 )} \left( w - \sqrt[L]{\lambda_{a}} , \overline{w} - \overline{\sqrt[L]{\lambda_{a}}} \right) \ra + \left( N_{\tot} - L N_{1} \right) \delta^{( 2 )} ( w , \overline{w} ) \right) =
$$
\begin{equation}\label{eq:EIG03}
= \frac{L N_{1}}{N_{\tot}} \left( \frac{1}{N_{1}} \sum_{a = 1}^{N_{1}} \la \delta^{( 2 )} \left( w - \sqrt[L]{\lambda_{a}} , \overline{w} - \overline{\sqrt[L]{\lambda_{a}}} \right) \ra \right) + \left( 1 - \frac{L N_{1}}{N_{\tot}} \right) \delta^{( 2 )} ( w , \overline{w} ) .
\end{equation}

Further, we recognize that the piece inside the big brackets in the lower line is precisely the mean density of the $L$--th roots of the eigenvalues of $\mathbf{P}$, hence, it will be equal to the distribution of these eigenvalues themselves upon the change of variables \smash{$w = \sqrt[L]{z}$},
$$
\frac{1}{N_{1}} \sum_{a = 1}^{N_{1}} \la \delta^{( 2 )} \left( w - \sqrt[L]{\lambda_{a}} , \overline{w} - \overline{\sqrt[L]{\lambda_{a}}} \right) \ra = \frac{1}{L} \frac{\partial z}{\partial w} \frac{\partial \overline{z}}{\partial \overline{w}} \rho_{\mathbf{P}} \left( w^{L} , \overline{w}^{L} \right) =
$$
\begin{equation}\label{eq:EIG04}
= L | w |^{2 ( L - 1 )} \rho_{\mathbf{P}} \left( w^{L} , \overline{w}^{L} \right) ,
\end{equation}
where the factor of $1 / L$ in the Jacobian originates from the $L$--valuedness of the transformation $z \to w$. Combining (\ref{eq:EIG04}) with (\ref{eq:EIG03}), we arrive at a relationship between the mean densities of the eigenvalues of \smash{$\widetilde{\mathbf{P}}$} and $\mathbf{P}$,
\begin{equation}\label{eq:EIG05}
\rho_{\widetilde{\mathbf{P}}} ( w , \overline{w} ) = \frac{L^{2} N_{1}}{N_{\tot}} | w |^{2 ( L - 1 )} \rho_{\mathbf{P}} \left( w^{L} , \overline{w}^{L} \right) + \left( 1 - \frac{L N_{1}}{N_{\tot}} \right) \delta^{( 2 )} ( w , \overline{w} ) .
\end{equation}
This generalizes formula (16) of~\cite{BurdaJanikWaclaw2010} to constituent random matrices having arbitrary rectangularities.

Relation (\ref{eq:EIG05}) may be recast (\ref{eq:MeanSpectralDensityFromNonHolomorphicGreensFunction}) in the language of the non--holomorphic Green's function,
\begin{equation}\label{eq:EIG06}
\frac{\partial}{\partial \overline{w}} G_{\widetilde{\mathbf{P}}} ( w , \overline{w} ) = \frac{\partial}{\partial \overline{w}} \left( \frac{L N_{1}}{N_{\tot}} w^{L - 1} G_{\mathbf{P}} \left( w^{L} , \overline{w}^{L} \right) + \left( 1 - \frac{L N_{1}}{N_{\tot}} \right) \frac{1}{w} \right) ,
\end{equation}
where in the last term on the RHS we have rewritten the complex Dirac's delta as \smash{$\delta^{( 2 )} ( w , \overline{w} ) = ( 1 / \pi ) \partial_{\overline{w}} ( 1 / w )$}, equivalent to (\ref{eq:ComplexDiracsDelta}) upon the regularization \smash{$1 / w = \lim_{\epsilon \to 0} \overline{w} / ( w \overline{w} + \epsilon^{2} )$}. It is of course desirable now to remove the derivatives \smash{$\partial_{\overline{w}}$} from both sides of (\ref{eq:EIG06}), but this could in principle produce an additive regular term independent of \smash{$\overline{w}$}, $f ( w )$; however, we will shortly justify that $f ( w ) = 0$. If so, (\ref{eq:EIG06}), without the derivatives, appears even simpler in terms of the non--holomorphic $M$--transform,
\begin{equation}\label{eq:EIG07}
M_{\widetilde{\mathbf{P}}} ( w , \overline{w} ) = \frac{L N_{1}}{N_{\tot}} M_{\mathbf{P}} \left( w^{L} , \overline{w}^{L} \right) ,
\end{equation}
which is exactly (\ref{eq:EIG07a}).

The idea of proving $f ( w ) = 0$ is to show that (\ref{eq:EIG07}) holds for the usual (holomorphic) $M$--transforms (\ref{eq:MTransformDefinition}), \ieNotAPPB outside of the domain $\mathcal{D}$ of the average eigenvalues; then, knowing that the non--holomorphic and holomorphic $M$--transforms coincide on the borderline $\partial \mathcal{D}$, (\ref{eq:EIG07}) would be proven. Hence, expanding the holomorphic $M$--transforms into the series of the moments (\ref{eq:MTransformExpansion}), and noticing that if an integer $n$ is a multiple of $L$, $n = L k$, then \smash{$\Tr \widetilde{\mathbf{P}}^{n} = L \Tr \mathbf{P}^{k}$} (\ref{eq:EIG02}), while in all the other cases, \smash{$\Tr \widetilde{\mathbf{P}}^{n} = 0$} --- we obtain
\begin{equation}\label{eq:EIG08}
M_{\widetilde{\mathbf{P}}} ( w ) = \sum_{n \geq 1} \frac{1}{w^{n}} \frac{1}{N_{\tot}} \Tr \la \widetilde{\mathbf{P}}^{n} \ra = \frac{L N_{1}}{N_{\tot}} \sum_{k \geq 1} \frac{1}{w^{L k}} \frac{1}{N_{1}} \Tr \la \mathbf{P}^{k} \ra = \frac{L N_{1}}{N_{\tot}} M_{\mathbf{P}} \left( w^{L} \right) ,
\end{equation}
\ieNotAPPB precisely like (\ref{eq:EIG07}), but in the holomorphic sector. This completes the proof of (\ref{eq:EIG07a}).


\section{Free Random Variables in a Nutshell}
\label{s:FreeRandomVariablesInANutshell}

In this appendix, we briefly outline some basic facts about the ``free random variables'' (FRV) calculus; given the broadness of the subject, we redirect the reader for more details to the specialized literature. The mathematical structure behind FRV, established by Voiculescu \emph{et al.} and Speicher~\cite{VoiculescuDykemaNica1992,Speicher1994}, is an extension of classical probability theory to the world of non--commuting random objects, of which large random matrices are a prime example.


\subsection{\ Freeness and Addition of Random Matrices}
\label{ss:FreenessAndAdditionOfRandomMatrices}


\subsubsection{\ Independence and the Classical Addition Law}
\label{sss:IndependenceAndTheClassicalAdditionLaw}

A fundamental notion in the standard theory of probability is that of independence --- two random variables \smash{$H_{1}$} and \smash{$H_{2}$} are called ``independent'' if after subtracting their average values, \smash{$H_{1 , 2}^{\prime} \equiv H_{1 , 2} - \langle H_{1 , 2} \rangle$}, they fulfil
\begin{equation}\label{eq:IndependenceDefinition}
\la H_{1}^{\prime} H_{2}^{\prime} \ra = 0 .
\end{equation}
Statistical independence constitutes a basis for numerous important constructions. For instance, the probability distribution of the sum \smash{$( H_{1} + H_{2} )$} of independent random numbers depends uniquely on the PDF's of the summands, and can be obtained by the following simple procedure: Thanks to (\ref{eq:IndependenceDefinition}), any moment of the sum can be expressed through the moments of the respective variables using the binomial theorem,
$$
M_{H_{1} + H_{2} , n} = \la \left( H_{1} + H_{2} \right)^{n} \ra = \sum_{k = 0}^{n} \binom{n}{k} \la H_{1}^{k} H_{2}^{n - k} \ra =
$$
\begin{equation}\label{eq:FactorizationOfClassicalMixedMoments}
= \sum_{k = 0}^{n} \binom{n}{k} \la H_{1}^{k} \ra \la H_{2}^{n - k} \ra = \sum_{k = 0}^{n} \binom{n}{k} M_{H_{1} , k} M_{H_{2} , n - k} .
\end{equation}
In other words, gathering the moments into a generating function, named the ``characteristic function,''
\begin{equation}\label{eq:CharacteristicFunctionDefinition}
g_{H} ( x ) \equiv \sum_{n \geq 0} \frac{\ii^{n} x^{n}}{n!} M_{H , n} = \la \sum_{n \geq 0} \frac{( \ii x H )^{n}}{n!} \ra = \la \ee^{\ii x H} \ra ,
\end{equation}
where $x$ is a real argument, one discovers that \smash{$g_{H_{1} + H_{2}} ( x ) = g_{H_{1}} ( x ) g_{H_{2}} ( x )$}, or yet in other words, that the logarithm of the characteristic function,
\begin{equation}\label{eq:LogarithmOfTheCharacteristicFunctionDefinition}
r_{H} ( x ) \equiv \log g_{H} ( x ) ,
\end{equation}
is simply additive when summing independent random numbers,
\begin{equation}\label{eq:ClassicalAdditionLaw}
r_{H_{1} + H_{2}} ( x ) = r_{H_{1}} ( x ) + r_{H_{2}} ( x ) , \quad \textrm{for independent \smash{$H_{1}$}, \smash{$H_{2}$}.}
\end{equation}
This is the ``commutative (classical) addition law.''


\subsubsection{\ Freeness}
\label{sss:Freeness}

One might wonder whether an analogous structure could be developed in the world of random matrices. Let us focus on the Hermitian case first --- consider $L$ Hermitian $N \times N$ random matrices \smash{$\mathbf{H}_{l}$}, and try to introduce an appropriate notion of independence. Let us also take the matrix dimension $N \to \infty$, which happens to be a necessary constraint: FRV works only in the thermodynamic limit.

One notices that it is not sufficient to assume statistical independence of all the entries of various \smash{$\mathbf{H}_{l}$}'s for them to be truly called ``independent'' --- indeed, one has to additionally require that there is no angular correlation between them; such correlations do arise from angular patterns specific to the given matrix ensembles. One may remove this angular dependence by performing a random similarity transformation of each matrix, with the uniform density on the unitary group, \smash{$\mathbf{U}_{l} \mathbf{H}_{l} \mathbf{U}_{l}^{\dagger}$} --- such matrices display then all the necessary features, and are referred to as ``freely independent,'' or just ``free.'' Non--commutative probability theory equipped with the notion of freeness is called ``free probability'' or ``free random variables'' (FRV) calculus.

In order to put this definition in rigorous terms~\cite{VoiculescuDykemaNica1992,Speicher1994}, one begins from making a claim that the appropriate generalization of the classical expectation map $\langle \ldots \rangle$ should be $\tau ( \ldots ) \equiv ( 1 / N ) \Tr \langle \ldots \rangle$. Now, one says that the \smash{$\mathbf{H}_{l}$}'s are free if their centered counterparts, \smash{$\mathbf{H}_{l}^{\prime} \equiv \mathbf{H}_{l} - \tau ( \mathbf{H}_{l} )$}, obey the following analogue of (\ref{eq:IndependenceDefinition}),
\begin{equation}\label{eq:FreenessDefinition}
\tau \left( \mathbf{H}_{l_{1}}^{\prime} \mathbf{H}_{l_{2}}^{\prime} \ldots \mathbf{H}_{l_{n}}^{\prime} \right) = 0 ,
\end{equation}
for all integers $n \geq 2$, and all indices \smash{$l_{1} , l_{2} , \ldots , l_{n} = 1 , 2 , \ldots , L$} such that \smash{$l_{1} \neq l_{2} \neq \ldots \neq l_{n}$}.

To get better acquainted with (\ref{eq:FreenessDefinition}), let us examine some of its simplest implications for two free matrices: By writing \smash{$\mathbf{H}_{l} = \mathbf{H}_{l}^{\prime} + \tau ( \mathbf{H}_{l} )$} and using (\ref{eq:FreenessDefinition}), one may work out all the correlation functions. The $2$--point one is then found to be
\begin{equation}\label{eq:TwoFreeRMTwoPointCorrelationFunction}
\tau \left( \mathbf{H}_{1} \mathbf{H}_{2} \right) = \tau \left( \mathbf{H}_{1} \right) \tau \left( \mathbf{H}_{2} \right) ,
\end{equation}
\ieNotAPPB precisely like in the commutative case. For the two possible $4$--point correlation functions, one has
\begin{equation}\label{eq:TwoFreeRMFourPointCorrelationFunction1}
\tau \left( \mathbf{H}_{1}^{2} \mathbf{H}_{2}^{2} \right) = \tau \left( \mathbf{H}_{1}^{2} \right) \tau \left( \mathbf{H}_{2}^{2} \right) ,
\end{equation}
\ieNotAPPB again no difference from the scalar sector, but
\begin{equation}\label{eq:TwoFreeRMFourPointCorrelationFunction2}
\tau \left( \mathbf{H}_{1} \mathbf{H}_{2} \mathbf{H}_{1} \mathbf{H}_{2} \right) = \tau \left( \mathbf{H}_{1} \right)^{2} \tau \left( \mathbf{H}_{2}^{2} \right) + \tau \left( \mathbf{H}_{1}^{2} \right) \tau \left( \mathbf{H}_{2} \right)^{2} - \tau \left( \mathbf{H}_{1} \right)^{2} \tau \left( \mathbf{H}_{2} \right)^{2} ,
\end{equation}
which is very different from the classical case. Hence, mixed moments generically do not factorize into moments of separate variables, as in usual probability theory --- a property which we have seen (\ref{eq:FactorizationOfClassicalMixedMoments}) to be the cornerstone of the classical addition law (\ref{eq:ClassicalAdditionLaw}); consequently, the latter will require some serious amendments to make up for this new situation.


\subsubsection{\ The FRV Addition Law for Hermitian Random Matrices}
\label{sss:TheFRVAdditionLawForHermitianRandomMatrices}

We have explained (\S\ref{sss:TheGreensFunctionAndMTransform}) that all the moments of a Hermitian random matrix $\mathbf{H}$ are conveniently grouped into a generating function such as the Green's function \smash{$G_{\mathbf{H}} ( z )$} (\ref{eq:GreensFunctionDefinition}) or the $M$--transform \smash{$M_{\mathbf{H}} ( z )$} (\ref{eq:MTransformDefinition}). In an analogous way as in standard probability theory, where the application of the logarithm (\ref{eq:LogarithmOfTheCharacteristicFunctionDefinition}) to the characteristic function \smash{$g_{H} ( x )$} (\ref{eq:CharacteristicFunctionDefinition}) leads to the object \smash{$r_{H} ( x )$}, which is additive (\ref{eq:ClassicalAdditionLaw}) under the addition of independent (\ref{eq:IndependenceDefinition}) random variables --- so in free probability, one has to apply \emph{functional inversion} to the Green's function,
\begin{equation}\label{eq:BluesFunctionDefinition}
G_{\mathbf{H}} \left( B_{\mathbf{H}} ( z ) \right) = B_{\mathbf{H}} \left( G_{\mathbf{H}} ( z ) \right) = z ,
\end{equation}
finding the so--called ``Blue's function,'' which fulfils the ``non--commutative (FRV) addition law,''
\begin{equation}\label{eq:NonCommutativeAdditionLaw}
B_{\mathbf{H}_{1} + \mathbf{H}_{2}} ( z ) = B_{\mathbf{H}_{1}} ( z ) + B_{\mathbf{H}_{2}} ( z ) - \frac{1}{z} , \quad \textrm{for free \smash{$\mathbf{H}_{1}$}, \smash{$\mathbf{H}_{2}$}.}
\end{equation}

Let us show on an example how this result can be practically used. Consider the normalized sum \smash{$( \mathbf{H}_{1} + \mathbf{H}_{2} + \ldots + \mathbf{H}_{L} ) / \sqrt{L}$} of $L$ free random matrices sampled from the Gaussian Unitary Ensembles (\ref{eq:GUEMeasure}), with the variances \smash{$\sigma_{1}^{2} , \sigma_{2}^{2} , \ldots , \sigma_{L}^{2}$}, respectively. The normalization by \smash{$1 / \sqrt{L}$} is equivalent to the rescaling of all the variances as \smash{$\sigma_{l}^{2} / L$}. The Green's function of the GUE with an arbitrary variance \smash{$\sigma^{2}$} is given by (\ref{eq:GreensFunctionForGUE}). It is easy to invert it functionally, which produces the corresponding Blue's function (\ref{eq:BluesFunctionDefinition}),
\begin{equation}\label{eq:BluesFunctionForGUE}
B_{\mathbf{H}} ( z ) = \sigma^{2} z + \frac{1}{z} .
\end{equation}
Hence, the addition law (\ref{eq:NonCommutativeAdditionLaw}) yields the Blue's function of the normalized sum to be
\begin{equation}\label{eq:BluesFunctionForTheSumOfLGUE}
B_{\left( \mathbf{H}_{1} + \mathbf{H}_{2} + \ldots + \mathbf{H}_{L} \right) / \sqrt{L}} ( z ) = \overline{\sigma^{2}} z + \frac{1}{z} , \quad \textrm{where} \quad \overline{\sigma^{2}} \equiv \frac{\sigma_{1}^{2} + \sigma_{2}^{2} + \ldots + \sigma_{L}^{2}}{L} .
\end{equation}
In other words, it remains a GUE random matrix, yet with the variance \smash{$\overline{\sigma^{2}}$} equal to the arithmetic mean of the constituent variances.

We have thus observed how one may exploit the FRV addition law (\ref{eq:NonCommutativeAdditionLaw}) to make discoveries of non--trivial results concerning sums of freely independent random matrices in an algorithmic, wholly algebraic, and usually quite simple way.

Let us finally remark that there exist several alternative notations commonly used in the literature, equivalent to the Blue's function language (\ref{eq:BluesFunctionDefinition}). For example, Voiculescu \emph{et al.} formulated the theory in terms of the ``$R$--transform,'' \smash{$R_{\mathbf{H}} ( z ) \equiv B_{\mathbf{H}} ( z ) - 1 / z$}, in terms of which the FRV addition law (\ref{eq:NonCommutativeAdditionLaw}) appears even simpler, \smash{$R_{\mathbf{H}_{1} + \mathbf{H}_{2}} ( z ) = R_{\mathbf{H}_{1}} ( z ) + R_{\mathbf{H}_{2}} ( z )$} (compare (\ref{eq:ClassicalAdditionLaw})).


\subsubsection{\ The FRV Addition Law for Non--Hermitian Random Matrices}
\label{sss:TheFRVAdditionLawForNonHermitianRandomMatrices}

Even though of no use in the current paper, let us mention that the FRV addition law (\ref{eq:NonCommutativeAdditionLaw}) can be generalized from the Hermitian to non--Hermitian realm with a ``quaternion construction''~\cite{JaroszNowak2004,JaroszNowak2006}.

The underlying idea looks as follows: We have elaborated (\S\ref{sss:TheNonHolomorphicGreensFunctionAndMTransform}) on how the $2 \times 2$ matrix--valued Green's function (\ref{eq:MatrixValuedGreensFunctionDefinition}) is the proper quantity to encode all the average spectral information about a non--Hermitian matrix model $\mathbf{X}$,
\begin{equation}\label{eq:MatrixValuedGreensFunctionDefinition2}
\mathcal{G}_{\mathbf{X}} ( z , \overline{z} ) = \lim_{\epsilon \to 0} \lim_{N \to \infty} \frac{1}{N} \bTr \la \left( \mathbf{Z}^{\DD}_{\epsilon} - \mathbf{X}^{\DD} \right)^{- 1} \ra .
\end{equation}
We have stated that it comprises a non--Hermitian analogue of the Hermitian Green's function \smash{$G_{\mathbf{H}} ( z )$} (\ref{eq:GreensFunctionDefinition}), but this is not entirely true --- the presence of \smash{$\mathbf{Z}^{\DD}_{\epsilon}$} suggests that it is rather a counterpart of \smash{$G_{\mathbf{H}} ( \lambda + \ii \epsilon )$}, \smash{$\epsilon \to 0^{+}$}. On the other hand, when one wishes to use the Hermitian FRV addition law (\ref{eq:NonCommutativeAdditionLaw}), one needs the Blue's function, which is the functional inversion of the Green's function (\ref{eq:BluesFunctionDefinition}), and hence, the functional form of the latter is necessary for any complex $z$, and not only for $z = \lambda + \ii \epsilon$.

In a pursuit of a non--Hermitian generalization of the Blue's function, one is therefore compelled to replace \smash{$\mathbf{Z}^{\DD}_{\epsilon}$} in (\ref{eq:MatrixValuedGreensFunctionDefinition2}) --- sufficient for producing the average spectral density, yet incapable of engineering functional inversion --- by an arbitrary \emph{quaternion},
\begin{equation}\label{eq:FromZDEpsilonToQ}
\mathbf{Z}^{\DD}_{\epsilon} \equiv \left( \begin{array}{cc} z & \ii \epsilon \\ \ii \epsilon & \overline{z} \end{array} \right) \otimes \Id_{N} \quad \longrightarrow \quad \mathcal{Q} \otimes \Id_{N} \equiv \left( \begin{array}{cc} a & \ii b \\ \ii \overline{b} & \overline{a} \end{array} \right) \otimes \Id_{N} ,
\end{equation}
where $a$, $b$ are complex numbers. (Actually, one might restrain $b$ to the real axis, \ieNotAPPB choose a $3$--dimensional subspace in the quaternion space. This, however, does not result in simplifying any equation, so we will abandon it. The point is that $b$ must no longer be close to zero, but cover the entire real line/complex plane.) In other words, the matrix--valued Green's function (\ref{eq:MatrixValuedGreensFunctionDefinition2}) is accordingly promoted to \emph{a quaternion function of a quaternion variable}, named the ``quaternion Green's function,''
\begin{equation}\label{eq:QuaternionGreensFunctionDefinition}
\mathcal{G}_{\mathbf{X}} ( \mathcal{Q} ) = \lim_{N \to \infty} \frac{1}{N} \bTr \la \left( \mathcal{Q} \otimes \Id_{N} - \mathbf{X}^{\DD} \right)^{- 1} \ra .
\end{equation}

After this modification, functional inversion is finally allowed in the quaternion space, and one defines the ``quaternion Blue's function,''
\begin{equation}\label{eq:QuaternionBluesFunctionDefinition}
\mathcal{G}_{\mathbf{X}} \left( \mathcal{B}_{\mathbf{X}} ( \mathcal{Q} ) \right) = \mathcal{B}_{\mathbf{X}} \left( \mathcal{G}_{\mathbf{X}} ( \mathcal{Q} ) \right) = \mathcal{Q} .
\end{equation}
Remarkably~\cite{JaroszNowak2004,JaroszNowak2006}, it may be proven that it satisfies the ``quaternion FRV addition law,'' which parallels the Hermitian version (\ref{eq:NonCommutativeAdditionLaw}), but at the quaternion level,
\begin{equation}\label{eq:QuaternionFRVAdditionLaw}
\mathcal{B}_{\mathbf{X}_{1} + \mathbf{X}_{2}} ( \mathcal{Q} ) = \mathcal{B}_{\mathbf{X}_{1}} ( \mathcal{Q} ) + \mathcal{B}_{\mathbf{X}_{2}} ( \mathcal{Q} ) - \frac{1}{\mathcal{Q}} , \quad \textrm{for free \smash{$\mathbf{X}_{1}$}, \smash{$\mathbf{X}_{2}$}.}
\end{equation}
This result constitutes a robust tool, greatly simplifying handling of sums of free non--Hermitian random matrices (see for instance~\cite{GorlichJarosz2004,GudowskaNowakJaroszNowakPapp2007} to view it at work).


\subsection{\ Multiplication of Random Matrices}
\label{ss:MultiplicationOfRandomMatrices}

Another important problem, after the addition of free random matrices, is their multiplication. For scalar random variables, it boils down to the addition task through the exponential change of variables, since \smash{$\exp H_{1} \exp H_{2} = \exp ( H_{1} + H_{2} )$}. Unfortunately, for non--commuting objects, generically \smash{$\exp \mathbf{H}_{1} \exp \mathbf{H}_{2} \neq \exp ( \mathbf{H}_{1} + \mathbf{H}_{2} )$}, and one faces a challenge to devise a different approach.

The FRV calculus provides such a straightforward prescription of multiplying free random matrices. They must be either Hermitian, but with an additional assumption that their product is Hermitian, too, or unitary, in which case the product automatically remains unitary. The procedure for such matrices looks as follows: First, invert functionally the pertinent $M$--transforms (\ref{eq:MTransformDefinition}),
\begin{equation}\label{eq:NTransformDefinition}
M_{\mathbf{H}} \left( N_{\mathbf{H}} ( z ) \right) = N_{\mathbf{H}} \left( M_{\mathbf{H}} ( z ) \right) = z ,
\end{equation}
obtaining the so--named ``$N$--transforms.'' This object is proven to satisfy the ``non--commutative (FRV) multiplication law,''
\begin{equation}\label{eq:NonCommutativeMultiplicationLaw}
N_{\mathbf{H}_{1} \mathbf{H}_{2}} ( z ) = \frac{z}{z + 1} N_{\mathbf{H}_{1}} ( z ) N_{\mathbf{H}_{2}} ( z ) , \quad \textrm{for free \smash{$\mathbf{H}_{1}$}, \smash{$\mathbf{H}_{2}$}.}
\end{equation}
We exploit this formula extensively in section~\ref{s:TheSingularValuesOfAProductOfRectangularGaussianRandomMatrices}.

Finally, let us mention that our notation is distinct from the one most often used in the context of multiplying free random variables; one may alternatively define the ``$S$--transform,'' \smash{$S_{\mathbf{H}} ( z ) \equiv ( z + 1 ) / ( z N_{\mathbf{H}} ( z ) )$}, which is simply multiplicative, \smash{$S_{\mathbf{H}_{1} \mathbf{H}_{2}} ( z ) = S_{\mathbf{H}_{1}} ( z ) S_{\mathbf{H}_{2}} ( z )$}.


\end{document}